\documentclass[12pt]{article}

\usepackage{color}

\usepackage{amsmath}
\usepackage{amssymb}
\usepackage{caption}
\usepackage{graphicx}
\usepackage{slashed}            % for slashed characters in math mode
\usepackage{subfig}             % for sub figures
\usepackage{xspace}				% For spacing after commands
\usepackage{cite}
\usepackage{float}
\usepackage{dcolumn}
\usepackage{bm}
\usepackage{feynmp}
\usepackage[english]{babel}
\usepackage{fancyhdr}
\usepackage{amsmath}
\usepackage{amssymb}
\usepackage{amsfonts}
\usepackage{psfrag}
\usepackage[applemac]{inputenc}
\usepackage{graphicx}
\newcommand{\drawsquare}[2]{\hbox{%
\rule{#2pt}{#1pt}\hskip-#2pt%  left vertical
\rule{#1pt}{#2pt}\hskip-#1pt%  lower horizontal
\rule[#1pt]{#1pt}{#2pt}}\rule[#1pt]{#2pt}{#2pt}\hskip-#2pt%  upper horizontal
\rule{#2pt}{#1pt}}% right vertical

\newcommand{\Yfund}{\drawsquare{7}{0.6}}%  fundamental
\newcommand{\Yafund}{\overline{\Yfund}}
\renewcommand{\tilde}{\widetilde} % dinky tildes look silly

\newcommand{\Tr}{\mathrm{Tr~}}

\newcommand{\TeV}{\,\mathrm{TeV}}

\newcommand{\beq}{\begin{eqnarray}}
\newcommand{\eeq}{\end{eqnarray}}

 \def\kahler{K\"ahler\ }

\newcommand{\bag}{\begin{align}}
\newcommand{\eag}{\end{align}}

\newcommand{\vev}[1]{\langle {#1} \rangle}

\usepackage[hypertexnames=false]{hyperref}		% kills links in index (if exists)

\usepackage{amsmath}%%%% PAGE DIMENSIONS
\usepackage{latexsym}

\usepackage{graphicx} % support the \includegraphics command and options

\begin{document}
\begin{titlepage}

\vskip.5cm

\begin{center} %TITLE HERE
{\huge \bf The $m_D-b_M$ Problem of Dirac Gauginos and its Solutions \vspace*{0.3cm} } 
\end{center}

\begin{center} % AUTHORS HERE
{\bf  Csaba Cs\'aki$^a$, Jessica Goodman$^b$,  Riccardo Pavesi$^a$, and Yuri Shirman$^c$} 
\end{center}
%\vskip 4pt

\begin{center} 
% PLACES HERE

$^{a}$ {\it Department of Physics, LEPP, Cornell University, Ithaca, NY 14853, USA} \\

\vspace*{0.1cm}

$^{b}$ {\it  Department of Physics,  The Ohio State University, Columbus, OH 43210} \\

\vspace*{0.1cm}

$^{c}$ {\it  Department of Physics,  University of California, Irvine, CA 92697} \\

\vspace*{0.1cm}

\vspace*{0.1cm}

{\tt  \href{mailto:csaki@cornell.edu}{csaki@cornell.edu}, 
 \href{mailto:jgoodman@physics.osu.edu}{jgoodman@physics.osu.edu},  
 \href{mailto:rp462@cornell.edu}{rp462@cornell.edu},
 \href{mailto:yshirman@uci.edu}{yshirman@uci.edu}}

\end{center}

\vglue 0.3truecm

\centerline{\large\bf Abstract}
\begin{quote}
 
 We examine the effective low-energy theory of the adjoint sector of Dirac gaugino models and its UV completions, and identify the main source of tuning. A holomorphic scalar adjoint mass square (the ``$b_M$ term") is generated at the same order (1-loop) as the Dirac gaugino mass (the ``$m_D$ term"), leading to the problematic relation $b_M\sim16\pi^2 m_D^2$, somewhat analogous to the $\mu-B_\mu$ problem of gauge mediation. We identify the leading operators of the low-energy effective theory contributing to the adjoint sector, and evaluate them in various UV completions, confirming the existence of this problem. We suggest a solution by introducing messenger mixing and tuning the relevant parameters. We also present a novel dynamical model for Dirac gauginos based on a strongly coupled SUSY QCD theory, where the additional adjoint $M$ is identified with a confined meson, the  $U(1)$ with a baryon-number like symmetry, and the messengers with the confined baryons. We find a SUSY breaking vacuum with a non-vanishing D-term, which after tuning the messenger mixing angles gives rise to a realistic gaugino and squark sector.
 
\end{quote}

\end{titlepage}

%\tableofcontents
%\newpage

\setcounter{equation}{0}
\setcounter{footnote}{0}
%%%%%%%%%%%%%%%%%%%%%%%%%%%%%%%%%%%%%%%%%%%%%%%%%%
%%%%%%%%%%%%%%%%%%%%%%%%%%%%%%%%%%%%%%%%%%%%%%%%%%
%%%%%%%%%%%%%%%%%%%%%%%%%%%%%%%%%%%%%%%%%%%%%%%%%%
%%%%%%%%%%%%%%%%%%%%%%%%%%%%%%%%%%%%%%%%%%%%%%%%%%
%%%%%%%%%%%%%%%%%%%%%%%%%%%%%%%%%%%%%%%%%%%%%%%%%%
%%%%%%%%%%%%%%%%%%%%%%%%%%%%%%%%%%%%%%%%%%%%%%%%%%
%%%%%%%%%%%%%%%%%%%%%%%%%%%%%%%%%%%%%%%%%%%%%%%%%%
%%%%%%%%%%%%%%%%%%%%%%%%%%%%%%%%%%%%%%%%%%%%%%%%%%

%%%%%%%%%%%%%%%%%%%%
%%%%%%%%%%%%%%%%%%%%
\section{Introduction}
\setcounter{equation}{0}
%%%%%%%%%%%%%%%%%%%%
%%%%%%%%%%%%%%%%%%%%

Supersymmetry (SUSY) has long been considered the best motivated candidate for physics beyond the Standard Model (SM) at the electroweak scale: it solves the hierarchy problem, predicts gauge coupling unification and provides a  viable dark matter candidate.  A dynamically broken SUSY theory can also offer a natural explanation for a large hierarchy between the SUSY breaking scale and the Grand Unified (GUT) or Planck scales. However, in the first three years of running the LHC and after 25 fb$^{-1}$ of data collected, no evidence for the existence of SUSY has been found. Instead, stringent bounds exceeding 1 TeV for the squark and gluino masses have been set by the ATLAS~\cite{AtlasSUSY} and CMS~\cite{CMSSUSY} experiments for the simplest SUSY models. The remaining possibilities  for avoiding collider bounds while preserving naturalness with SUSY  include light stops and heavy first and second generations~\cite{moreminimal,naturalsusy,Brust:2011tb}, a stealthy spectrum~\cite{stealth}, R-parity violation~\cite{RPV} and Dirac gluinos~\cite{Fox:2002bu}. 

In this paper we consider the scenario with Dirac gauginos in detail. The idea of a Dirac-type gluino mass was first proposed by Fayet in the 70's~\cite{Fayet}. The main motivation for such models  (besides it simply being an interesting possibility different from the canonical Majorana gluino case \cite{Hall:1990hq}) is that there are several improvements to naturalness: the stop contribution to the Higgs mass is now finite and cut off by the Dirac gluino mass~\cite{Fox:2002bu}; the splitting between gluinos and squarks can be increased \cite{Fox:2002bu, Brust:2011tb}; and the squark production cross sections are somewhat suppressed as same-handedness t-channel gluino exchange requires a chirality flipping Majorana mass and is therefore absent, while at the same time mixed-handedness processes are suppressed by the second power of the gluino mass rather than just the first~\cite{Kribs:2012gx}. Furthermore, flavor constraints can be significantly alleviated when R-symmetries are present~\cite{Kribs:2007ac}.
Finally, the presence of R-symmetries is also natural in many models of dynamical SUSY breaking.
For other aspects of R-symmetric models with Dirac gaugino masses see \cite{Goodsell,Benakli, Amigo:2008rc, Abel:2011dc, Belanger:2009wf,Blechman:2009if, Harnik:2008uu, Hsieh:2007wq,McCullough,Kribs:2010md}.

The first part of this paper is dedicated to a careful analysis of the low-energy effective theory in Dirac gaugino models. There are three soft breaking parameters relevant for the adjoint sector: the Dirac mass $m_D$, a holomorphic adjoint scalar mass $b_M$ and a real scalar mass $m_M^2$. We will show that a realistic mass spectrum would require all of these parameters to be roughly of the same order $m_D^2\sim b_M \sim m_M^2$. However, the two holomorphic terms (the Dirac gaugino mass and the holomorphic scalar mass) generally appear at the same order in a small expansion parameter. For example, in messenger models considered in this paper both of these terms are generated
at one loop, leading to the problematic relation $b_M \sim 16 \pi^2 m_D^2$, similar to the $\mu-B_\mu$ problem~\cite{dvali,muBmu} in gauge mediation. We will refer to this as the ``$m_D-b_M$ problem" of Dirac gluinos. After identifying this problem we set out to calculate the operators contributing to these parameters in a variety of models with increasing structure among the messengers. The simplest supersoft model with unmixed messengers indeed suffers from the $m_D-b_M$ problem: $b_M$ and $m_D$ are generated at one loop, while $m_M^2$ at two loops. We identify the effective operators contributing to $m_M^2$, and argue that in order to capture the contributions proportional to $D$ or $D^2$ one needs to include the dynamical fields $\psi, \bar{\psi}$ giving rise to the D-term of the theory. To actually find a realistic spectrum, one needs to include mixing among the messenger fields: by tuning the mixing parameters one can find regions where the $b_M$ term is suppressed compared to its natural size, while there will be additional one-loop contributions to $m_M^2$, also suppressed by mixing parameters.

In the second half of the paper we set out to find a working dynamical example for generating a viable Dirac gaugino sector. The simplest way to achieve this is to identify the adjoints with a meson of a confining theory, and the additional $U(1)$ needed for generating the D-term with a baryon-number like symmetry of the strong sector. The necessary messenger-adjoint interaction is then automatically generated, while the SM gauge group can be embedded into a global $SU(5)$ symmetry of the model. We show what  additional fields and superpotential terms are needed in order to obtain a realistic vacuum. After a careful examination of the vacuum structure of the theory, we set out to investigate the $m_D-b_M$ problem in an explicit example. As expected from the general arguments we find that for generic parameters the adjoint sector is not realistic (one of the scalar mass squares is negative), however by tuning the messenger mixings small regions of parameter space with a realistic adjoint sector can be identified.
This conforms with the generic expectation of an order $1/{16\pi^2} \sim 0.1-1\%$ tuning.
Once this tuning is performed, a realistic gaugino and squark sector can be obtained.
However, as we explain in our brief discussion of the phenomenology, additional complications arise once we also take the slepton and Higgs sector into account, which will require an extension of the model to raise these masses.

The paper is organized as follows. In Sec.~\ref{problem} we present the general form of the $m_D-b_M$ problem. In Sections \ref{secholomorphic} and \ref{kahler} we perform the operator analysis of Dirac gaugino models, Sec.~\ref{secholomorphic} focusing on the holomorphic operators, while Sec.~\ref{kahler} discusses the real operators and the effects of messenger mixing. Sec.~\ref{model} contains the setup for the dynamical model. In Sec.~\ref{vacuum} we examine the vacuum structure of the model. We discuss the tuning necessary to obtain a realistic adjoint scalar sector in Sec.~\ref{adjointmasses}, and in Sec.~\ref{secpheno} we comment on the basic features of the phenomenology. We conclude in Sec.~\ref{conclusion}, and an Appendix contains an alternative mechanism for generating realistic adjoint scalar masses using discrete symmetries.

%%%%%%%%%%%%%%%%%%%%
%%%%%%%%%%%%%%%%%%%%
\section{Dirac Gauginos and the $m_D-b_M$ Problem \label{problem}}
\setcounter{equation}{0}
%%%%%%%%%%%%%%%%%%%%
%%%%%%%%%%%%%%%%%%%%

A supersymmetric extension of the SM with Dirac gauginos contains an adjoint chiral superfield $M_i, i=1,2,3$ for every gauge group, whose fermionic component pairs up with the gauginos $\lambda_i$, while the scalars become massive. This adjoint sector  is controled by the  the Lagrangian
\begin{equation}
{\cal L}_{eff} \supset -\left( m_D \lambda \psi_M + b_M M^2 +{\rm h.c.} \right) - m_M^2 |M|^2\ ,
\label{eq:effecLan}
\end{equation}
where $m_D$ is the Dirac mass parameter, $b_M$ is the holomorphic adjoint scalar mass, while $m_M$ is a real mass for the adjoint scalar. The mass eigenvalues of the two real scalar fields living in the adjoint supermultiplet $M$ are given by\footnote{For simplicity we will assume that $b_M$ is real and positive, thus giving a positive mass squared to the real component of $M$ and possibly negative mass squared to the imaginary component of $M$.} $\delta_{\pm}^2=m_M^2\pm 2 |b_M|$. 
To ensure that the SM gauge symmetries are not broken via adjoints, we clearly need $m_M^2 \geq 2|b_M|$. The dangerous possibility of obtaining a negative adjoint mass square has been noticed in many previous papers, see e.g.~\cite{Fayet,Fox:2002bu,Goodsell}. 

The MSSM sfermion masses (the masses of the scalar partners of the Standard Model fermions) are generated through loops of gauge bosons, gauginos, and the scalar components of the adjoint $M$, calculated in~\cite{Fox:2002bu}. These diagrams are finite and give contributions to the sfermion masses of the form
\begin{equation}
m^2_r=\frac{C_i(r) \alpha_i m_D^2}{\pi}\text{log} \Big( \frac{\delta^2_+}{m_D^2} \Big) \ .
\label{eq:sfermionmass}
\end{equation}
However, loops of the scalar adjoint give another negative contribution to the  sfermion masses, which has been emphasized recently in~\cite{Stanfordstudents}. If the supertrace of the adjoint sector (the chiral adjoint and vector supermultiplets) is non-vanishing, it 
feeds into the RGE equation for the MSSM sfermions at two loops making them run from the scale $\Lambda$ where the adjoint masses are generated and generically lowering their values. 
Alternatively, one can capture adjoint contributions to the running of the squark masses by treating them as messengers in the usual gauge mediation (albeit in a different representation and with non-vanishing supertrace).
The $SU(2)_L$ triplet scalar adjoint also contributes, but the larger color gauge coupling will dominate.  The two-loop RGE has been first analyzed in~\cite{Goodsell:2012fm}, 
and the coefficient was corrected in~\cite{Stanfordstudents} using the general expression from~\cite{Martin:1993zk}. The resulting RGE is:
\begin{equation}
\frac{d m_{Q}^2}{dt}=\frac{16}{(16 \pi^2)^2}g_3^4 {\rm STr} M^2_8
\label{eq:twoloop}
\end{equation}
which also follows directly from the original two-loop calculation of scalar masses in \cite{Poppitz:1996xw}, where the adjoint $M$ is treated as a messenger.  In this equation ${\rm STr} M^2_8$ is the supertrace over the octet sector. Note that the supertrace appears because holomorphic operators will contribute both to $m_D$ and to $m_M^2$ but their effect in (\ref{eq:twoloop}) will cancel. From (\ref{eq:effecLan}) $\mathrm{STr} M^2_8=2m_M^2-4m_D^2$. If $m_M$ is too large, some of the squark mass squares will become negative, thus one obtains a bound of the form~\footnote{Here and in the rest of the paper we do not distinguish between $m_M^2$ and $\mathrm{STr}M_8^2$. It is, however, possible to construct models where $\mathrm{STr}M_8^2=0$ and (\ref{eq:mMbound}) does not apply. See the Appendix for an example.}
\begin{equation}
\frac{m_M^2}{m_D^2} \lesssim \frac{\pi C}{\alpha} \frac{\log \frac{\delta}{m_D}}{\log \frac{\Lambda}{m_Q}} \sim {\cal O}(10)\ ,
\label{eq:mMbound}
\end{equation}
where $\Lambda$ is the scale at which the adjoint scalar masses are generated, while $\delta$ is the scalar adjoint mass. For concreteness we will be imposing 
  $\frac{m_M^2}{m_D^2}\leq 10$  in accordance with the detailed analysis performed in~\cite{Stanfordstudents}.

To obtain a realistic spectrum, one would like $m_D \sim 1-3 $ TeV, which combined with the bound (\ref{eq:mMbound}) implies $m_M< 3-10$ TeV to avoid negative mass squares in the MSSM sfermion sector. To also avoid a negative mass square among the adjoint scalars we will need $b_M \sim {\cal O}(m_M^2)$. Thus a realistic spectrum can only be achieved if $m_D^2 \sim b_M \sim m_M^2$.\footnote{$b_M \ll m_D^2 \sim m_M^2$ would also be acceptable.}

The relation  $m_D^2 \sim b_M \sim m_M^2$ leads to the fine tuning problem of Dirac gaugino models.
Both $m_D$ and $b_M$ are {\em holomorphic} soft terms arising from chiral (superpotential) operators in the
superfield Lagrangian. Thus,  
one may expect that they are generated by the same dynamics and, in terms of a small expansion parameter $\epsilon$,
$m_D\propto \epsilon, b_M \propto \epsilon$. On the other hand, the magnitude of $m_M^2$ is model dependent. If SUSY breaking arises from the superpotential operators,
the supertrace of the masses must vanish implying that $m_M^2\sim m_D^2\sim \epsilon^2$ (as we will see below).\footnote{To see this consider the one-loop effective potential: the supertrace of the masses is the coefficient of the quadratic divergence. Chiral operators only renormalize \kahler potential terms which, by power counting, can only include log-divergent terms or inverse powers, no quadratic divergences. So the supertrace has to vanish.} On the other hand, $m_M^2$ may also receive additional contributions from \kahler potential operators. While such contributions are negligible in minimal UV completions of supersoft models of \cite{Fox:2002bu}, they could also be very large in non-minimal models.

Regardless of the magnitude of $m_M^2$, the relation $b_M \gg m_D^2$, violates the scaling relations necessary for a realistic spectrum.  This is somewhat similar to the $\mu -B_\mu$ problem~\cite{dvali,muBmu} in ordinary gauge mediation, where both $\mu$ and $B_\mu$ are generated at one loop, leading to $B_\mu \gg \mu^2$, while a realistic spectrum requires the two to be of the same order.\footnote{In this analogy $m_M^2$ is similar to a non-holomorphic soft Higgs mass, $m_{H}^2$, which is determined by the mediation mechanism and  generically unrelated to the $\mu$-term.} In the same vein we can refer to this hierarchy as the $m_D-b_M$ problem of Dirac gaugino models. In the following we will investigate what the natural values of these parameters $m_D,b_M$ and $m_M^2$ are in some simple dynamical models, and show how one can eventually arrive at a model where the parameters can be tuned (at the order $\epsilon$ level) to satisfy the required hierarchies.

\section{Holomorphic Operators in the Adjoint Sector\label{secholomorphic}}
\setcounter{equation}{0}

In order to find the parameters $m_D, b_M$ and $m_M^2$ introduced above, we need to identify the effective operators contributing to these various mass terms. First, we will focus on the holomorphic operators, and in the next section we will investigate terms arising from the K\"ahler potential.

\subsection{The supersoft operator for the Dirac mass}

To obtain Dirac gaugino masses, one needs a  D-term SUSY breaking in a hidden sector $U(1)$ group in addition to the adjoint fields $M_j$ for each standard model gauge group, as explained by Fox, Nelson and Weiner in~\cite{Fox:2002bu}.  The D-term breaking can be parametrized by introducing the field strength $W_\alpha$ of the hidden $U(1)$ symmetry with
\begin{equation}
W_\alpha = D \theta_\alpha +\ldots\ .
\end{equation}
The Dirac gaugino masses are generated by the effective holomorphic operator 
\begin{equation}
W_{\text{ssoft}} = e^{i\alpha_D} \frac{W_\alpha W^\alpha_j M_j}{\Lambda}.
\label{eq:ssoft}
\end{equation}
where  $W^\alpha_j$ are the standard model field strengths, $\Lambda$ is the cutoff corresponding to some physical mass scale in the UV completion of the theory, and the phase $\alpha_D$ is determined by the underlying dynamics.  This operator was dubbed the supersoft operator in  \cite{Fox:2002bu} because it only introduces finite corrections to the masses of the superpartners of the standard model fields, instead of the logarithmically divergent ones in standard soft breaking. Once SUSY breaking is introduced through the D-term of the $U(1)$ vector multiplet,  this operator will contain a Dirac mass term of the form $\sim \frac{D}{\Lambda} \lambda_j \psi_{M_j}$, partnering gauginos to the fermions in $M$. Setting the phase $\alpha_D=0$ the component Lagrangian now contains the following terms:
\begin{equation}
 \mathcal{L}\supset-\frac{D}{\Lambda}\lambda_i\psi_{M_i}-\frac{D^2}{\Lambda^2}(M_i+M_i^\dagger)^2
-\sqrt{2}\frac{D}{\Lambda}(M_i+M_i^\dagger)\left(\sum_j g_i q_j^\dagger t_i q_j\right)
\end{equation}
We see that the operator in  (\ref{eq:ssoft}) results in a Dirac gaugino mass given by $m_D=\frac{D}{\Lambda}$. In addition, it also contributes to the parameters $b_M$ and $m_M^2$ by the amounts $m_M^2=2b_M = \frac{2D^2}{\Lambda^2}$, leading to a contribution to the real part of the scalar adjoint mass equal to twice the Dirac mass, while the imaginary part remains massless. 

\begin{figure}[H]
\centering{
 \includegraphics[width=.375\textwidth]{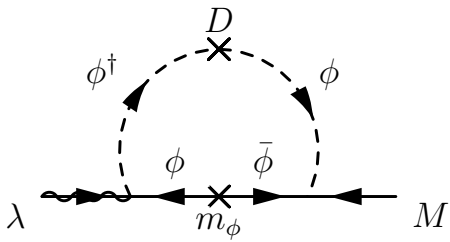}
}
\caption{The one-loop diagram leading to the supersoft operator via integrating out the heavy messengers.}
\label{fig:GauginoLoop}
\end{figure}

 A simple way~\cite{Fox:2002bu} of generating the supersoft operator (\ref{eq:ssoft}) is to introduce a coupling of heavy messengers, $\phi$ and $\bar{\phi}$  charged under the $U(1)$ symmetry to the adjoint superfield $M$ via  the superpotential 
\begin{equation}
W = y \bar{\phi} M\phi + m_{\phi}\bar{\phi}\phi \ .
\label{eq:SSGen}
\end{equation}
The supersoft operator will then be generated after integrating out the messengers as shown in the diagram in Fig.~\ref{fig:GauginoLoop} which contributes 
\begin{equation}
m_{Di} = \frac{y g_i g D}{16 \pi^2 \sqrt{2}m_\phi} \text{; } b_M=m_D^2 \text{; } m_M^2=2|m_D|^2
\label{eq:gauginoloop}
\end{equation}
where $g_i$ is the SM gauge coupling and $g$ is the hidden $U(1)$ gauge coupling. 
Note that $ \alpha_D=\mathrm{arg} (y)-\mathrm{arg}(m_\phi)$, $\mathrm{arg}(m_D)=\alpha_D$, and $\mathrm{arg}(b_M)=2\alpha_D$. Furthermore, one scalar in the adjoint multiplet remains massless independently of this phase.

\subsection{Holomorphic scalar adjoint mass \label{holomorphic}}

It was noted in \cite{Fox:2002bu} that one more supersoft operator consistent with all gauge and global symmetries can be added to the superpotential:
\begin{equation}
 \int d^2\theta e^{i\alpha_b}\frac{W_\alpha W^\alpha}{\Lambda^2}M_i^2+{\rm h.c.} \, ,
\label{eq:ssoftbadguy}
\end{equation}
where once again the phase $\alpha_b$ is determined by the underlying dynamics. This operator contributes directly to $b_M$ but not to $m_M^2$. For $\alpha_b=0$ we find
\begin{equation}
 \mathcal{L}\supset-\frac{D^2}{\Lambda^2} (M_i^2+{M_i^\dagger}^2)\,.
\end{equation}
In general, if
$\alpha_b=2\alpha_D$, the two contributions to $b_M$ from (\ref{eq:ssoft}) and (\ref{eq:ssoftbadguy}) have the same phase. Therefore, in the absence of additional contributions to $m_M^2$ the mass squared matrix of 
the adjoint $M$ will have one negative eigenvalue.
Indeed, in the simplest model considered above with messengers $\phi ,\bar\phi$ coupled to the adjoint, this term will be generated at one loop, and its coefficient can be estimated easily without actually performing a loop calculation. 
The reason is that this term can be related to the running of the hidden $U(1)$ coupling constant.  
Integrating out the messengers at $m_\phi$ we find that  the coefficient of the $W_\alpha^2$ term in the effective low energy description is
\begin{equation}
-\left( \frac{1}{4g^2(\Lambda)} +\frac{b_L}{16\pi^2} \log \frac{\mu}{m_\phi} +\frac{b_H}{16\pi^2} \log \frac{m_\phi}{\Lambda} \right) W_\alpha W^\alpha\ ,
\end{equation}
where $b_{H,L}$ are the $U(1)$ $\beta$-function coefficients above and below the $m_\phi$ threshold. In the minimal model $b_L-b_H=5q^2$ where $q$ is the $U(1)$ charge of $\phi$ and $\bar\phi$ and the factor of 5 arises since messengers transform in a fundamental representation of $SU(5)_{\mathrm{GUT}}$. The messenger mass $m_\phi$ can be thought of as the VEV of the meson $M$ and the formula above generalizes to 
\begin{equation}
-\left( \frac{1}{4g^2(\Lambda)} +\frac{b_L}{16\pi^2} \log \frac{\mu}{\Lambda} -\frac{q^2}{16\pi^2} \log \frac{\det{(m_\phi+ y M)}}{\Lambda^5} \right) W_\alpha W^\alpha\,.
\end{equation}

\begin{figure}[H]
\centering{
 \includegraphics[width=.8\textwidth]{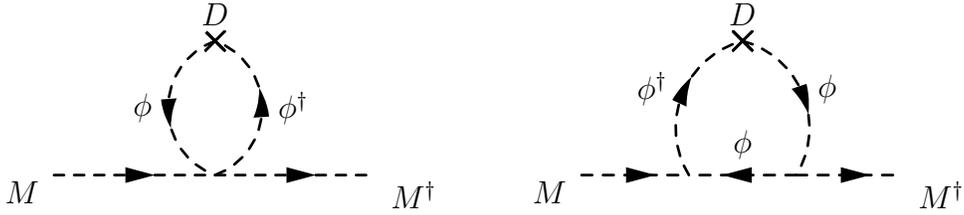}
}
\caption{\label{fig:linearD} The diagrams contributing to the $MM^\dagger$ mass term at leading order in $D$. There are two other diagrams with $\phi \to \bar{\phi}$ which will have an extra sign due to the sign of D, and on the cubic diagram the insertion can also be done on the lower line.}
\end{figure}

By expanding this expression in powers of $y M$ we find that the full expression of the operator (\ref{eq:ssoftbadguy}) is given by 
\begin{equation}
-\int d^2\theta \frac{q^2 y^2 g^2}{32\pi^2} \frac{1}{m_\phi^2} W_\alpha W^\alpha \Tr M^2 +{\rm h.c.}
\label{eq:holM2mass}
\end{equation}
The resulting $b_M$ term is given by $b_M= \frac{q^2 y^2 g^2 D^2}{32 \pi^2 m_\phi^2}$ and its phase is $\mathrm{arg}(b_M)\equiv \alpha_b=2\alpha_D$.
Thus the simplest UV completion of the supersoft model has one scalar with negative mass squared.

Independently of the signs of scalar masses, we
can appreciate the difficulty of the $m_D-b_M$ problem already at this stage. Both $m_D$ and $b_M$ are generated at one loop, giving rise to a hierarchy $b_M \sim 16\pi^2 m_D^2$. On the other hand, there is no symmetry which can forbid this term (\ref{eq:ssoftbadguy})  if the original supersoft operator for the Dirac mass (\ref{eq:ssoft}) is allowed, just like the case of the $B_\mu$ term. 

%%%%%%%%%%%%%%%%%%%%
%%%%%%%%%%%%%%%%%%%%
\section{K\"ahler Operators for the Scalar Adjoint Sector \label{kahler}}
\setcounter{equation}{0}
%%%%%%%%%%%%%%%%%%%%
%%%%%%%%%%%%%%%%%%%%

We have seen above that the two holomorphic operators (\ref{eq:ssoft}) and (\ref{eq:ssoftbadguy}) 
cannot yield a realistic model on their own and additional terms giving  positive contributions to $m_M^2$ are required.  It was suggested in~\cite{Carpenter:2010as} that such terms are indeed generated at order $D^2$, corresponding to the operator 
\begin{equation}
 \int d^4 \theta \frac{W_\alpha D^\alpha V}{\Lambda^2}M^\dagger M\, .
\label{eq:Linda}
\end{equation}
This was argued to happen at one loop order in a model 
where the spontaneous breaking of the $U(1)$ gauge symmetry leads to the generation of a non-vanishing $D$-term at the minimum of the potential. However, the fact that this term is not $U(1)$ gauge invariant\footnote{We thank Graham Kribs for bringing this to our attention.} indicates that (\ref{eq:Linda}) is not actually the right form for a K\"ahler term contributing to $m_M^2$. 
Below, we reconsider models with spontaneous $U(1)$ breaking, find the correct operators, the order at which they appear in perturbation theory, and explain how to take the limit in which the minimal supersoft scenario is recovered.

\begin{figure}[H]
\centering{
 \includegraphics[width=.75\textwidth]{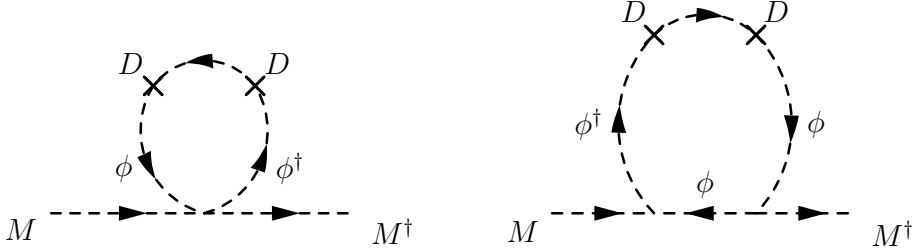}
}
\caption{\label{fig:quadraticD} The diagrams contributing to the $MM^\dagger$ mass term at quadratic order in $D$. There are two other diagrams with $\phi \to \bar{\phi}$, and on the cubic diagram the insertion can also be done on the lower line, or one on the upper and one on the lower line, giving rise to the factor of 3 in (\ref{eq:quadraticDcancel}).}
\end{figure}

\subsection{Real adjoint mass in the minimal supersoft model}

To verify our expectation that (\ref{eq:Linda}) is not actually the right effective operator to capture a real adjoint mass we first consider a model with an explicit Fayet-Illiopoulos term in the Lagrangian (in other words we assume that the mass of the $U(1)$ breaking field is much larger than its VEV so that the $U(1)$ is effectively unbroken). 
We calculated the scalar adjoint $MM^\dagger$ mass terms both linear and quadratic in $D$ and have found that both of those vanish in the simplest supersoft model with a single (or several unmixed) messengers. 
The diagrams contributing to the $MM^\dagger$ mass term at linear order in $D$ are given in Fig.~\ref{fig:linearD}, and the coefficient of the contribution from $\phi$ is given by 
\begin{equation}
y^2\int p^2 d p^2\left[ \frac{1}{(p^2-m_\phi^2)^2} -\frac{2m_\phi^2}{(p^2-m_\phi^2)^3}\right] 
\end{equation}
Once we add the $\bar{\phi}$ contribution we find that the two cancel.
Similarly, the contributions to the $D^2$ term in $MM^\dagger$ are given in Fig.~\ref{fig:quadraticD}. While there are no cancellations between diagrams with $\phi$ and $\bar\phi$ fields, they vanish individually since the coefficient is now proportional to 
\begin{equation}
y^2\int p^2 d p^2\left[ \frac{1}{(p^2-m_\phi^2)^3} -\frac{3m_\phi^2}{(p^2-m_\phi^2)^4}\right] =0\,.
\label{eq:quadraticDcancel}
\end{equation}
This cancellation is reminiscent of the accidental cancellation of $\mathcal{O}(F^2)$ terms in GMSB models with messenger-matter mixing \cite{dvali,Dine:1996xk}. Similarly to GMSB models, the subleading $\mathcal{O}(D^4)$ terms will be non-vanishing, but they are negligible for our purposes. On the other hand, in contrast to models of \cite{dvali,Dine:1996xk}, the absence of gauge invariant operators implies that leading order terms will not be generated even at higher loop orders in perturbation theory.

The cancellations we see here  imply that in order to generate additional direct contributions to $m_M^2$ (and to solve the problem with the negative scalar adjoint masses) one needs to consider models with additional ingredients. 

\subsection{Dynamical fields for U(1) breaking - is supersoft ever supersoft?}

Before considering our full model with messenger mixing, let us first analyze an intermediate case, where one has charged messengers $\phi, \bar{\phi}$ as before, but in addition we introduce the singlets $\psi, \bar{\psi}$ that break the U(1) symmetry and give rise to the non-vanishing D-term
\begin{equation}
D \propto |\psi|^2 -|\bar{\psi}|^2\ .
\label{eq:psiDterm}
\end{equation}
In this case, a new gauge invariant contribution to the scalar adjoint masses can be written using the fields $\psi , \bar{\psi}$:
\begin{equation}
 \int d^4 \theta \frac{1}{\Lambda^2}\left[\psi ^\dagger e^{q V}\psi+\bar\psi ^\dagger e^{-q V}\bar\psi\right] {\rm Tr} M^\dagger M\,.
\label{eq:Yuri}
\end{equation}
This operator is manifestly gauge invariant and we will argue that it is generated at one loop in the presence of Yukawas while it is always generated at two loops through $U(1)$ gauge interactions. 
To prove this last statement, we can interpret this operator as the usual $U(1)$ gauge mediated soft SUSY breaking mass term for the fields $\psi$ and $\bar{\psi}$, where now $M$ is assumed to be the spurion with an F-term. We assume that $\psi$ and $\bar{\psi}$ are light and calculate their wavefunction renormalization in the infrared as a function of $M$. This will contain the operator in (\ref{eq:Yuri}). If $M$ had an F-term this would generate a non-holomorphic soft mass for the light fields, and hence the standard GMSB result can be applied to calculate its coefficient.
The reason is that the fields $\phi$ and $\bar{\phi}$ can be treated as 5 independent messenger pairs for $U(1)$ gauge mediation as in Sec.~\ref{holomorphic} and the standard method of analytic continuation in superspace \cite{Giudice:1997ni} can be used to extract the coefficient of the operator in (\ref{eq:Yuri}) from expanding the one-loop wavefunction renormalization in each of the 5 thresholds.
The result is analogous to the simplest one-threshold case, differing only in producing a trace for the matrix of spurions:
\begin{equation}
 -\int d^4 \theta  \frac{1}{m_\phi^2}\left(\frac{\alpha(\mu)q_\phi q}{2 \pi}\right)^2 \left[\psi ^\dagger e^{q V}\psi+\bar\psi ^\dagger e^{-q V}\bar\psi\right] {\rm Tr} M^\dagger M\,.
\end{equation}
It is important to remember that we are interested in the limit where the $U(1)$ Higgs fields $\psi$ and $\bar\psi$ are at least as massive as the $U(1)$ gauge field. Thus, the coefficient of (\ref{eq:Yuri}) becomes a function of both $m_\phi$ and $m_\psi$. In the limit $m_\psi\gg \vev{\psi}$ the coefficient of (\ref{eq:Yuri}) vanishes recovering the result obtained in the previous subsection.
 
Once the operator in (\ref{eq:Yuri}) is generated, non-supersoft contributions to the soft masses of the MSSM sfermions will be generated \footnote{To see this is not supersoft: the spurion is $m_M^2 \theta^4$ and the operator in \ref{eq:nonss} giving sfermion soft masses can be constructed from this without additional cutoff dependence}
\begin{equation}
 \int d^4\theta \frac{1}{\Lambda^2} \left[\psi ^\dagger e^{q V}\psi+\bar\psi ^\dagger e^{-q V}\bar\psi\right]  Q^\dagger Q \,.
\label{eq:nonss}
\end{equation}
While non-supersoft contributions to sfermion masses are suppressed by an additional loop factor, they are also enhanced by large logs due to RGE's and may become important unless SUSY is broken at low scales.

Before moving on to non-minimal models, it is worth noting that while the operator (\ref{eq:Yuri}) is formally linear in $D$, it is also proportional to $|\psi|^2-|\bar\psi|^2$. In the minimal model this proportionality leads to  an additional factor of $D$. Thus on-shell (\ref{eq:Yuri}) generates soft terms of the type anticipated in~\cite{Carpenter:2010as} albeit with a different coefficient and at higher order in the perturbative expansion.

\subsection{A realistic adjoint mass spectrum via messenger mixing \label{withmixing}}
\label{sec:realistic}

We are now finally ready to consider the third toy example, where the $U(1)$ is dynamical as in the previous case, however in addition there is mixing between the messengers $\phi ,\bar{\phi}$ and additional messengers $N,\bar{N}$ not charged under $U(1)$, described by the following superpotential:

\begin{equation}
 W=y(\bar\phi M\phi + \bar\psi \phi \bar N+\psi \bar\phi N)+m_\phi \bar\phi \phi\, .
 \label{eq:phiNmessangers}
\end{equation}
  We will again assume that due to some other dynamics $\psi$ and $\bar \psi$ develop VEVs and generate a small $D$-term as in (\ref{eq:psiDterm}).
The main new ingredient is the mixing of the messengers which can resolve the negative mass square problem: as long as $\langle \psi \rangle \ne \langle \bar\psi \rangle$, which we have assumed in order to generate a nonzero $D$-term,  the mixing between $\phi$ and $N$ and $\bar\phi$ and $\bar N$ will differ and the contributions from $\phi$ will no longer cancel those from $\bar\phi$ in Fig.~\ref{fig:linearD}. 
While these contributions are formally $\mathcal{O}(D)$, they are further suppressed by the difference in mixing angles and can naturally be comparable to other soft terms.
Furthermore, the calculation of $\mathcal{O}(D^2)$ diagrams in Fig.~\ref{fig:quadraticD} will also be modified: diagrams with cubic and quartic vertices will have different number of mixing angle insertions and the cancellation we saw earlier will not persist.\footnote{It is amusing to note that once again the situation is reminiscent of a disappearance of a similar cancellation in GMSB models with messenger-matter mixing in the presence of several spurions \cite{Intriligator:2010be}.}

Thus, mixing will result in new contributions to both $MM^\dagger$ and $MM$ scalar masses.
In particular, the operator in (\ref{eq:Yuri}) will actually be generated at one loop (rather then the generic two loop as argued before), which will give rise to contributions to $m_M^2$ of order    
\begin{equation}
 m_M^2\sim \frac{y^2}{16\pi^2}\frac{\left( |\psi|^2 -|\bar{\psi}|^2\right) g D}{m_\phi^2}\, ,
\end{equation}
where we have not included the mixing factors.  In models where the $U(1)$ is broken by VEVs of several Higgses while only $\psi$ and $\bar\psi$ contribute to messenger mixing, it is possible to obtain any desired value of $gD/(|\psi|^2-|\bar\psi|^2)$. Therefore one could construct models where the $\mathcal{O}(D)$ and $\mathcal{O}(D^2)$ contributions to the soft adjoint masses are comparable,  while higher order in $D$ contributions are small 
due to $\frac{D}{m_\phi^2}$ $\ll1$. Indeed, this scenario will be realized in the model we discuss in the upcoming sections. 

The effects of the messenger mixings can be roughly estimated in the following way. We will denote the magnitude of the mixing of the $\phi$ messenger by $c_1$ (with $c_1=1$ corresponding to the case of no mixing) and the mixing of $\bar\phi$ by $c_2$. In the limit when the VEVs of $\psi$ and $\bar\psi$ are equal one expects $c_1=c_2$. There will now be heavy and light fields running in the loops of Figs.~\ref{fig:linearD}-\ref{fig:quadraticD}. 
The $\phi -\bar\phi$ fermion propagator 
will pick up a factor of $c_1 c_2$ for the heavy fields (and $s_1 s_2$ for the light), the scalar $\phi \phi^*$ propagator will pick up mixing factors of $c_1^2$ (and $s_1^2$) for the heavy (light) fields, and the scalar $\bar \phi\bar\phi^*$ propagator will pickup factors $c_2^2$ ($s_2^2$) for heavy (light) fields.
Based on this, the expected magnitudes of the various mass parameters are displayed in Table \ref{fig:masstable}. The gaugino mass just picks up an ${\cal O}(1)$ correction due to the mixing. The holomorphic $b_M$ parameter had a cancellation between the $\phi$ and $\bar\phi$ contribution to the term linear in $D$. After mixing this cancellation is no longer expected to occur. $m_M^2$ had two separate cancellations: one between $\phi$ and $\bar\phi$ in Fig.~\ref{fig:linearD} and a separate one between the quartic and the cubic terms in Fig.~\ref{fig:quadraticD}. None of those are expected to remain in the presence of generic mixing angles.

\begin{table}
\begin{center}
\[ \begin{array}{|c|c|c|}
\hline & {\rm w/o\  mixing}  & {\rm with\ mixing} \\ \hline 
\vphantom{\frac{x^{x^{x}}}{x^{x^{X}}}} m_D & \frac{y g_s}{16\pi^2} \frac{D}{m_\phi} & \frac{y g_s}{16\pi^2} \frac{D}{m_\phi} c_1 c_2 (c_1^2+c_2^2) \\ \hline
\vphantom{\frac{x^{x^{x}}}{x^{x^{X}}}} b_M & \frac{y^2}{16\pi^2} \frac{D^2}{m_\phi^2} & \frac{y^2}{16\pi^2} (c_1^6-c_2^6) D + \frac{y^2}{16\pi^2} (c_1^8+c_2^8) \frac{D^2}{m_\phi^2} \\ \hline
\vphantom{\frac{x^{x^{x}}}{x^{x^{X}}}} m_M^2 & {\cal O} \left( \frac{1}{16\pi^2} \right)^2 & \frac{y^2}{16\pi^2} \left[ c_1^4(1+c_1^2) -c_2^4 (1+c_2^2) \right] D +\frac{y^2}{16\pi^2} \left[ c_1^6(1-c_1^2) +c_2^6 (1-c_2^2) \right]\frac{D^2}{m_\phi^2} \\ \hline 
\end{array} \]
\end{center}
\caption{Estimates of the magnitudes of the various mass parameters in the absence of messenger mixing (left column), and in the presence of mixing (right column).  \label{fig:masstable}}
\end{table}

Now we can see how to resolve the $b_M-m_D$ problem and obtain a realistic adjoint mass spectrum.\footnote{The possibility that messenger mixing might result in a realistic adjoint sector was also investigated in~\cite{Goodsell}.} 
 The simplest way is by tuning the amount of mixings among the messengers such that the formally one-loop $b_M$ is suppressed to be numerically of two-loop size. This suggests that a tuning between two terms of order $1/16\pi^2$ will be necessary: as in most realistic SUSY models a complete Dirac gaugino model will inevitably contain a percent level tuning among the parameters, unless a natural way of suppressing the $b_M$ term can be obtained (however as we mentioned no symmetry can be used for that). Here we will just assume that the messenger mixings can be adjusted to obtain a realistic spectrum. 
The necessary tuning to achieve a realistic spectrum is:
\begin{itemize}
\item The mixing should be small ($c\sim1$) to suppress $m_M^2$ and satisfy the bound (\ref{eq:mMbound}).
\item The terms linear in $D$ vs. quadratic in $D$ in the expression of $b_M$ have to conspire to give a term that is much smaller than the individual terms: the sum should be about a loop factor smaller than the terms themselves.
\end{itemize}
Once that tuning is achieved all mass parameters will be about the same magnitude, and a realistic Dirac gaugino model will result. Another possible solution to the $m_D-b_M$-problem using discrete symmetries is exhibited in App.~\ref{App:discrete}.

%%%%%%%%%%%%%%%%%%%%
%%%%%%%%%%%%%%%%%%%%
\section{Dynamical Dirac Gauginos - The Setup \label{model}}
\setcounter{equation}{0}
%%%%%%%%%%%%%%%%%%%%
%%%%%%%%%%%%%%%%%%%%

The aim of the second half of the paper is to present a concrete dynamical realization of the ideas explored so far: a model where a SUSY breaking D-term appears dynamically, together with the additional adjoint coupled to messengers as in (\ref{eq:SSGen}). The low energy dynamics of strongly interacting SUSY QCD theories \cite{Seiberg:1994bz} generically includes a composite meson field coupled to either dual quarks or baryons. It is then natural to try to identify the meson with the additional adjoint $M$, and the messengers with components of the dual quarks/baryons (see also \cite{Amigo:2008rc}). In this case the SM gauge group can be embedded into the flavor symmetry of low energy SQCD. Thus, the flavor symmetry should at least be $SU(5)$. However, to obtain a SUSY breaking vacuum, at least one composite field should get a VEV, hence the minimal size of the flavor symmetry in the DSB sector is $SU(6)$. To focus on the simplest possibility we assume that there is no gauge group left in the magnetic description though this is not essential. Therefore we will take as our starting setup $SU(5)$ SUSY QCD with $6$ flavors similar to the model in \cite{Csaki:2006wi}.  This theory is s-confining \cite{Seiberg:1994bz,Csaki:1996sm}, $(F=N+1)$ and below the strong coupling scale $\Lambda$ the appropriate description is given in terms of the gauge invariant mesons ($\tilde{M} \sim \frac{1}{\Lambda}\bar{Q}Q$), baryons ($B \sim \frac{1}{\Lambda^4}Q^5$) and anti-baryons ($\bar{B} \sim \frac{1}{\Lambda^4}\bar{Q}^5$).  
The UV theory contains the following fields: 
\begin{equation} \nonumber
\begin{array}{c| c |c c c  c}
&SU(5) & SU(6) & SU(6) & U(1)_B & U(1)_R \\ \hline 
\vphantom{\frac{x^{x^{x}}}{x^{x^{X}}}}  Q & \Yfund & \Yafund & \mathbf{1} & 1 & \frac{1}{6} \\
\vphantom{\frac{x^{x^{x}}}{x^{x^{X}}}} \bar{Q} &\Yafund & \mathbf{1} & \Yfund & -1 & \frac{1}{6} \\ \hline
\end{array}
\end{equation}

The charges of the low-energy degrees of freedom are given by
\begin{equation} 
\begin{array}{c| c c c c}
&SU(6) & SU(6) & U(1)_B & U(1)_R \\ \hline 
\vphantom{\frac{x^{x^{x}}}{x^{x^{X}}}} \tilde{M} & \Yafund& \Yfund & 0 & \frac{1}{3} \\
\vphantom{\frac{x^{x^{x}}}{x^{x^{X}}}} B & \Yfund & \mathbf{1} & 5 & \frac{5}{6} \\
\vphantom{\frac{x^{x^{x}}}{x^{x^{X}}}} \bar{B} & \mathbf{1} & \Yafund & -5 & \frac{5}{6} \\ \hline
\end{array}
\label{eq:magfields}
\end{equation}
together with the superpotential, $W=y(\bar{B}\tilde{M}B-\frac{1}{\Lambda^3}\rm{det}\tilde{M})$.
We give a mass to one flavor, $W_{tree}=m_{66}Q_{6}\bar{Q}_6$ reducing the $SU(6)$ flavor symmetry.  We weakly gauge the diagonal $SU(5)$ subgroup of the remaining flavor symmetry allowing us to embed the Standard Model gauge group in it.   In particular, we will parametrize the mesons, baryons, and anti-baryons as follows
$$ \tilde{M}=  \left( \begin{matrix}
M&N\\\bar{N}&X
\end{matrix}\right), \, B=\left( \begin{matrix} \phi \\ \psi \end{matrix}\right), \, \bar{B}=\left( \begin{matrix} \overline{\phi} \\ \overline{\psi} \end{matrix}\right),$$
where $X,\psi,\bar{\psi}$ are now $SU(5)$ singlets, $N\text{ and }\phi$ are $SU(5)$ fundamentals, $\bar{N}\text{ and }\bar{\phi}$ $SU(5)$ anti-fundamentals, and $M$ is an adjoint plus a singlet under $SU(5)$.  The low-energy superpotential is then given by
\begin{equation}
W_1 = y(\bar{\phi} M \phi + \psi N \bar{\phi} + \bar{\psi}  \bar{N} \phi+\psi \bar{\psi} X)-\mu^2X
\label{eq:W1}
\end{equation}
where the first four terms originate from $\bar{B}\tilde{M}B$ and the last term corresponds to $W_{tree}$. As long as the VEV's of the fields are $\ll\Lambda$, the instanton generated $\frac{1}{\Lambda^3}{\rm det}\tilde{M}$ can be neglected. However, it will re-introduce SUSY preserving minima far from the origin, making the minimum close to the origin metastable as in \cite{Intriligator:2006dd}.

If the meson $\mathrm{Tr}M$ develops a VEV, the first three terms in the superpotential (\ref{eq:W1}) correspond to the superpotential (\ref{eq:phiNmessangers}) making the identification between the low energy degrees of freedom in the dynamical models and fields in the model of Sec. \ref{sec:realistic} trivial.
The Yukawa coupling in (\ref{eq:SSGen}) between the adjoint and the messengers will then be generated via the strong dynamics in (\ref{eq:W1}). Recall from (\ref{eq:SSGen}) that the messengers also require a supersymmetric mass to generate (\ref{eq:ssoft}). Thus we must give Tr$M$ a VEV to generate this mass term. This can be achieved by adding a coupling to an additional singlet $\chi$, which will also make Tr$M$ massive.
Finally, the $U(1)$ symmetry of the superpotential (\ref{eq:phiNmessangers}) required for generating Dirac gaugino masses can be identified with the baryon number of the SUSY QCD model (\ref{eq:magfields}). 

While D-term supersymmetry breaking by itself is impossible to obtain dynamically \cite{Dumitrescu:2010ca}, it has been shown to emerge naturally under generic conditions together with a SUSY-breaking F-term of comparable magnitude. Thus, we want to perturb the SQCD model (\ref{eq:magfields}) in a way that does not destroy its local SUSY breaking minimum while simultaneously generating a non-vanishing D-term. This D-term must arise from asymmetric VEVs of the $SU(5)$ singlet components of the baryons and anti-baryons ($\psi$ and $\bar{\psi}$).  To force $\langle\psi\rangle\neq\langle\bar{\psi}\rangle$, we add additional singlet fields $S,\bar{S},T,\bar{T}$, charged under $U(1)_B$,$\text{ and }Z$ with superpotential interactions of the form 
\begin{equation}
W_2= h S (\psi+T)+h'\bar{S}(\bar{\psi} +\bar{T})+\alpha Z T\bar{T}
\end{equation}
It is interesting to note that for some choices of the $U(1)_B$ gauge coupling charge conjugation may be broken spontaneously and a non-vanishing D-term exists even in the limit where $h = h'$.  Note that we could also add a term linear in $Z$ since $Z$ and $X$ have the same charges. However, as long as the corresponding parameter is smaller than $\mu^2$ we will obtain the same vacuum structure.

The  $S\psi$ and $\bar{S}\bar{\psi}$ terms are important for obtaining the desired vacuum. They arise from dimension 6 superpotential operators in the UV theory and are expected to be suppressed by three powers of the cutoff in the IR  $h\sim \Lambda (\frac{\Lambda}{M_{UV}})^3$.
This is the only deformation by an irrelevant operator that we include. While there are other operators consistent with our symmetries that have lower dimension in the UV, for example $ST^2\bar{T}$, they remain irrelevant in the IR and their contributions at the minimum are negligible.

We will see that in order to obtain phenomenologically interesting results we will have to be able to vary the amount of mixing between the fields $\phi$, $\bar{\phi}$, $N \text{ and } \bar{N}$. To achieve this we must also introduce the fields $N'$ and $\bar{N}'$ with interactions $W_3=m_{N'} N \bar{N}'+m_{N'} \bar{N} N'$.
Combining all the ingredients, the full matter content and symmetries of the model in the IR are given in Table \ref{tabfields}, with the superpotential 
\begin{eqnarray}
W = W_1+W_2+W_3+m' \chi (\text{Tr}M-v_{M})
\label{eq:fullW}
\end{eqnarray}

\begin{table}
\begin{equation} \nonumber
\begin{array}{c|  c c c c c c c c c| c c c c c c c c }

& M &\text{Tr}M & N &\bar{N} & X & \psi & \bar{\psi} & \phi & \bar{\phi} & S & \bar{S} & T & \bar{T} 
& Z & \chi & N' & \bar{N'} \\ \hline 
\vphantom{\frac{x^{x^{x}}}{x^{x^{X}}}} SU(5) & \text{Adj} & \mathbf{1} &\Yfund&\Yafund &\mathbf{1}&\mathbf{1}&\mathbf{1}&\Yfund&\Yafund&\mathbf{1}&\mathbf{1}&\mathbf{1}&\mathbf{1}&\mathbf{1}&\mathbf{1}&\Yfund&\Yafund \\
\vphantom{\frac{x^{x^{x}}}{x^{x^{X}}}} U(1)_{\text{B}}&0&0&0&0&0&1&-1&1&-1&-1&1&1&-1&0&0&0&0 \\
\vphantom{\frac{x^{x^{x}}}{x^{x^{X}}}} U(1)_{\text{R}}&0&0&1&1&2&0&0&1&1&2&2&0&0&2&2&1&1\\ \hline
\end{array} 
\label{eqn:IRFields}
\end{equation}
\caption{The matter content of the full dynamical model together with the charges under the relevant symmetry}
\label{tabfields}
\end{table}

%%%%%%%%%%%%%%%%%%%%
%%%%%%%%%%%%%%%%%%%%
\section{The Vacuum of the Dynamical Model \label{vacuum}}
\setcounter{equation}{0}
%%%%%%%%%%%%%%%%%%%%
%%%%%%%%%%%%%%%%%%%%

To find the vacua of the model outlined in Section \ref{model} we have numerically minimized the scalar potential obtained from (\ref{eq:fullW}). We found non-zero F-terms for $X,S,\bar{S}, \text{ and }Z$, while  in addition to Tr$M$ the fields $\psi, \bar{\psi}, T, \text{ and }\bar{T}$ will also obtain unequal VEVs. These VEVs provide the non-zero D-term needed for generating Dirac gaugino masses through the supersoft operator. Thus the vacuum will be characterized by the following VEVs:
\begin{eqnarray}
\langle D\rangle \ne 0 \\
\langle F_X\rangle, \langle F_S\rangle,\langle F_{\bar{S}}\rangle, \langle F_Z\rangle \ne 0 \\
\langle \psi \rangle,\langle \bar{\psi} \rangle, \langle T\rangle, \langle \bar{T} \rangle \ne 0\\
\langle \text{Tr}M \rangle = v_{M}
\end{eqnarray}
The structure of the minimum can be understood in the following heuristic way. First we note that the F-terms for $\bar{N}'$ and $N'$ force to zero the VEVs of $N$ and $\bar{N}$.  Next we consider $F_{\phi}$ and $F_{\bar{\phi}}$, which set the VEVs for $\phi$ and $\bar{\phi}$ to zero.  Finally, the F-terms for $N$ and $\bar{N}$ force the VEVs for $\bar{N}'$ and $N'$ to zero while the F-term for $M$ forces $\langle \chi \rangle = 0$.  This leaves F-terms for $X,\psi, \bar{\psi}, T,\bar{T}, S, \bar{S}$ and $Z$ as well as the $U(1)$ D-term to be minimized.  Allowing $hS=-y \bar{\psi}X$, $h'\bar{S}=-y \psi X$, and $\alpha Z T = y \psi X$ further sets $F_{\psi}$, $F_{\bar{\psi}}$, and $F_{\bar{T}}$ to zero.  The remaining scalar potential along this slice is
\begin{eqnarray}
V &=& |y \psi \bar{\psi} - \mu^2|^2 
+|y \psi X(\frac{\bar{\psi}}{\psi}-\frac{\bar{T}}{T})|^2+|h(\psi+T)|^2+ |h'(\bar{\psi}+\bar{T})|^2+|\alpha T\bar{T}|^2\nonumber \\
&+&\frac{1}{2}g^2\left((1+|\frac{y X}{h'}|^2)|\psi|^2-(1+|\frac{y X}{h}|^2)|\bar{\psi}|^2+|T|^2-|\bar{T}|^2\right)^2
\end{eqnarray}
The F-term for $T$ can be set to zero by allowing $X=0$ or $\frac{\bar{\psi}}{\psi}=\frac{\bar{T}}{T}$.  However, $F_S$ and $F_{\bar{S}}$ want $T \sim \psi$ and $\bar{T} \sim \bar{\psi}$ while $F_Z$ wants $T\bar{T} \sim 0$ and $F_X$ wants $\psi \bar{\psi} \sim \mu^2$.  Because of this, $X=0$ allows for a lower energy therefore also setting $S,\bar{S},Z=0$ and eliminating the flat direction usually associated with non-zero F-terms in O'Raifeartaigh models. Essentially the F-flat direction has a built in asymmetry between charge conjugate fields, however this direction cannot be D-flat. Thus the only classical flat direction in the model is $M$ adjoint, which as we will see, is lifted by loops.

While there may be several interesting regions in the parameter space, in the following discussion we will focus on one of them:
\begin{equation}
\begin{aligned}
&\mu \sim 0.5-5 \times 10^4 \rm{TeV}\\
&\Lambda \gtrsim 1000\mu \\
&v_ {M}\sim 10^4-10^5 \rm{TeV}
\end{aligned}
\end{equation}
\noindent $\mu$ sets the overall scale of the model, and is chosen such that the final resulting gaugino masses are in the few TeV range. $v_{M}$ sets the messenger masses. While the lower bound on $\Lambda$ follows from the requirement that the ${\rm det} M$ term be negligible.
Values of $h$ in the range $h\sim h' \sim$(0.5-5)$\mu$ are most interesting since for smaller or larger values the D-term is very small compared to the F term. This implies that a coincidence of scales of the form $\mu\sim\Lambda (\frac{\Lambda}{M_{UV}})^3$ has to occur in this model imposed by the requirement of having a significant amount of D-term breaking. This is the only coincidence needed for the presence of an interesting vacuum.

\begin{figure}[H]
\centering{
\includegraphics[width=.45\textwidth]{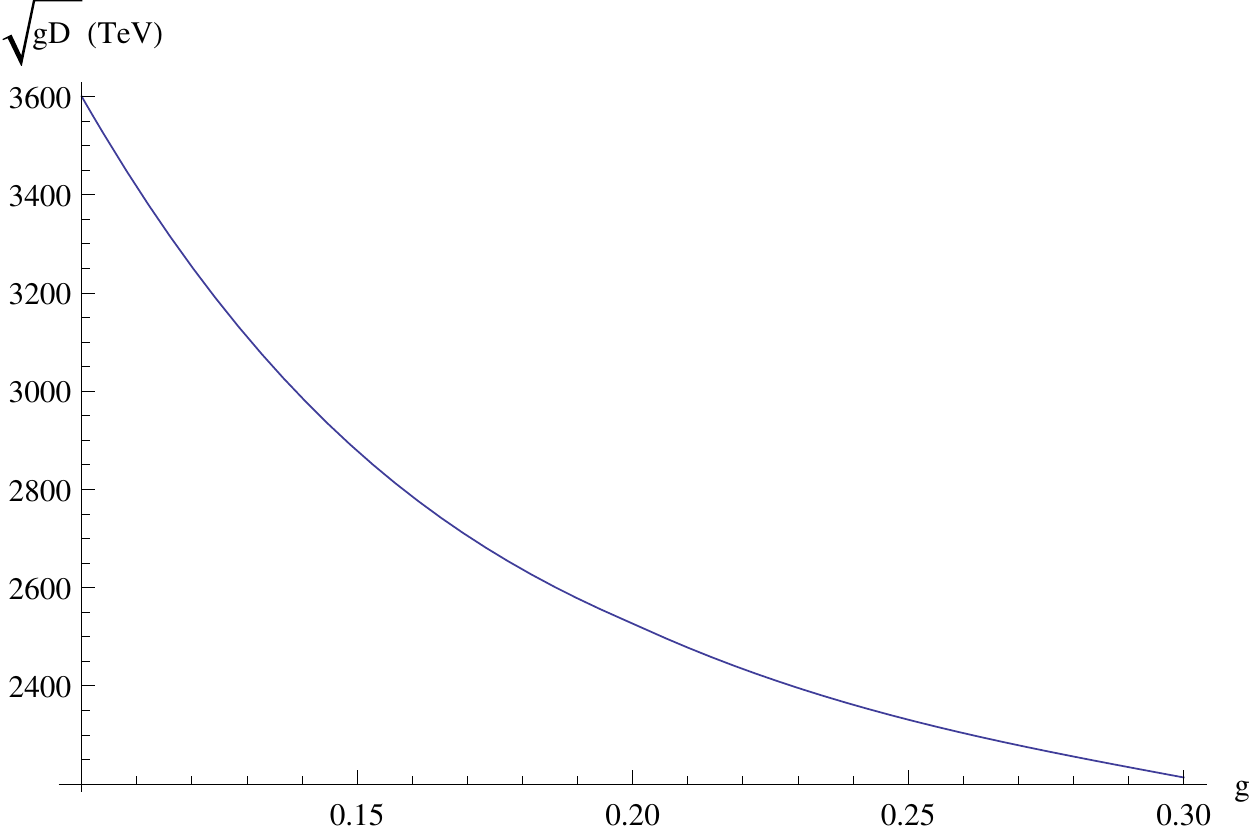}\hspace*{1.5cm}\includegraphics[width=.45\textwidth]{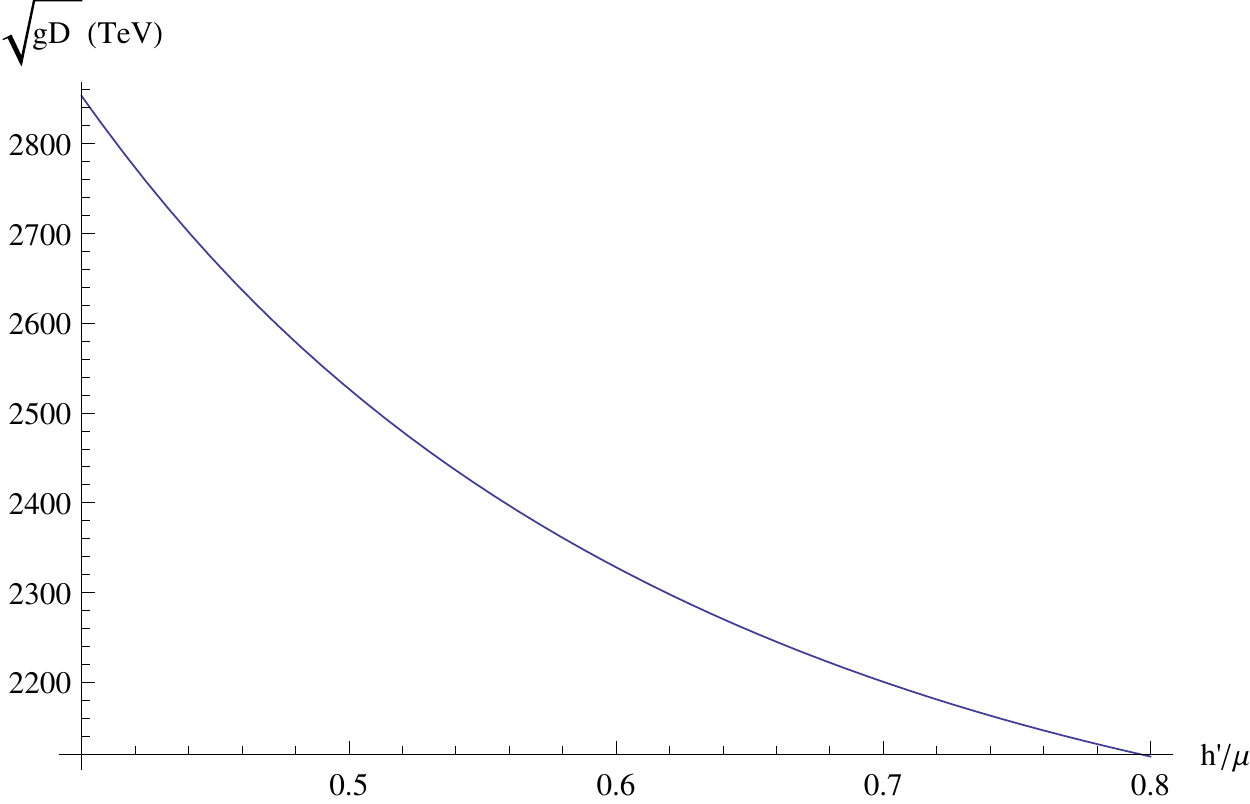}\hspace*{2cm}
}
\caption{$\sqrt{gD}$ versus $g$ for $h'$=0.5$\mu$ (left) and vs $h'$ for $g=0.2$ (right). In both cases we fixed $h=5\mu$ and $\alpha$=1.7, while the overall scale is set to obtain 2.5 TeV Dirac gluino masses}
\label{fig:Dplot}
\end{figure}

The combination that is relevant for the loop generated masses is $g D$, where $g$ is the $U(1)$ gauge coupling and $D$ is the $U(1)$ D-term and so we plot this as a function of the relevant parameters in Fig.~\ref{fig:Dplot}. The non-vanishing D-term is a measure of the asymmetry in the VEVs, and all the masses of the Dirac gaugino model are set by it. The plots show that our metastable minimum is always present and keeps its features for a wide range of parameters.

\noindent A representative example of VEV's for the fields is:
\begin{equation}
\psi \sim \mu \;\;
\bar{\psi}\sim 0.85\mu \;\;
T \sim -\mu \;\;
\bar{T} \sim -0.07\mu
\end{equation}

\noindent for $h=5\mu$, $h'=0.5\mu$, $\alpha$=1.7, $g$=0.2 and $m_{N'}$=1.15$\mu$. The VEVs never differ from $\mu$ by more than a factor of few, with one larger and one smaller than $\mu$ for each pair $\psi$,$\bar{\psi}$ and $T$, $\bar{T}$. One of the fields from $T$ and $\bar{T}$ is getting a much smaller VEV (typically at least an order of magnitude smaller, depending on the relative sizes of $g$ and $\alpha$).

\begin{figure}[H]
\centering{
 \includegraphics[width=.5\textwidth]{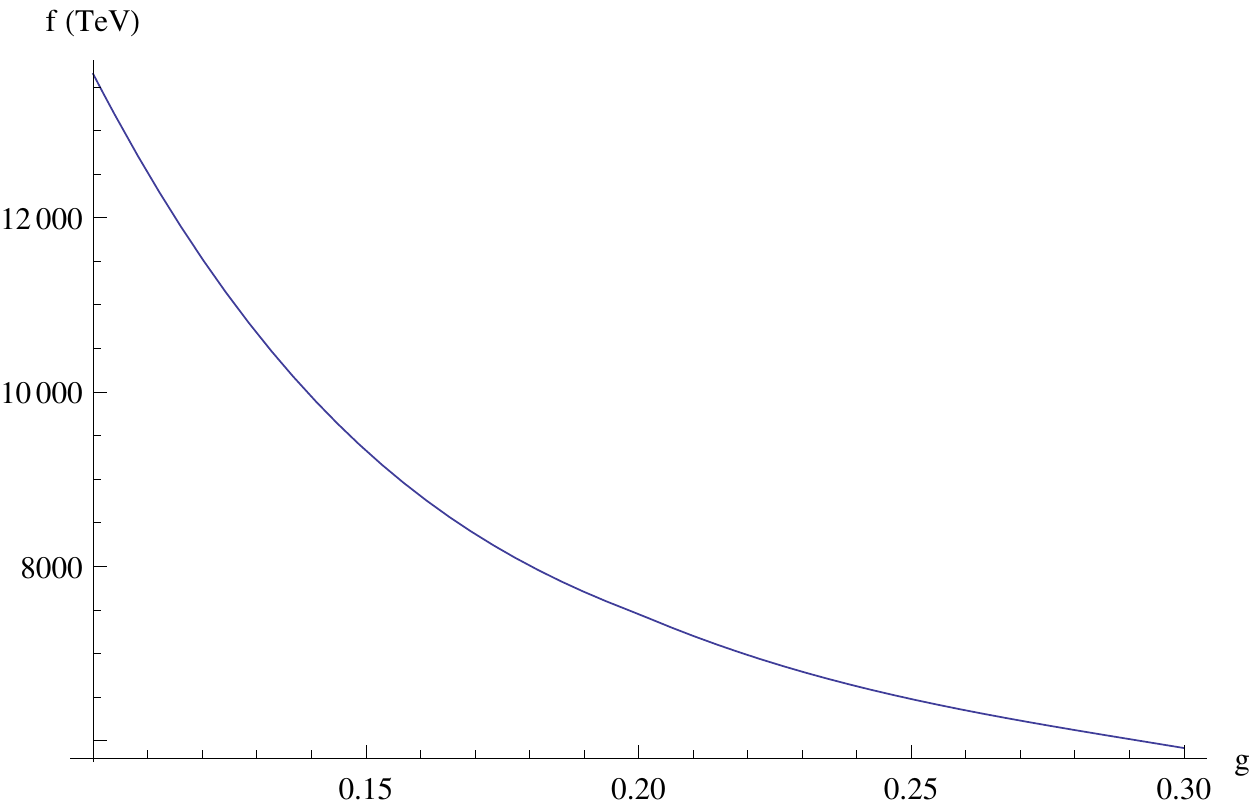}
}
\caption{The scale of SUSY breaking versus $g$, for $\alpha$=1.7, $v_{M}\sim 5\times 10^4$ TeV. Note that we are fixing the Dirac gluino mass to 2.5 TeV.}
\label{fig:scale}
\end{figure}

The magnitude of SUSY breaking can be quantified by $f=(F^2+\frac{1}{2}D^2)^{1/4}$, which corresponds to the value of the scalar potential at the minimum and sets the scale of the gravitino mass. This quantity is quite insensitive to the dimensionless parameters, since the role of $g$ and $\alpha$ is to control the size of D which is subdominant to F.
In Figure \ref{fig:scale} we show the scale of SUSY breaking, as a function of $g$.  Note that the insensitivity to $g$ is not visible in the plots since we are holding the gluino mass fixed at 2.5 TeV and thus different values of $g$ correspond to varying $\mu$ together with the scale of the model as well.

\section{Scalar adjoint spectrum in the dynamical model\label{adjointmasses}}

The dynamical model presented in Sec.\ref{model} and expanded around the vacuum explored in Sec.\ref{vacuum} provides an explicit implementation of the models presented in Sec.\ref{kahler}, where we explained that mixing among the messengers can be used to obtain a realistic mass spectrum. The model in \ref{eq:fullW} without the $N', \bar{N'}$ fields corresponds exactly to the one investigated in Eq.(\ref{eq:phiNmessangers}), with the identification $m_\phi=\frac{v_{M}}{5}$. However, in this case both the Dirac gaugino mass (due to the chirality flipping fermion propagator in Fig.~\ref{fig:GauginoLoop}) as well as the messenger mixings are controlled by the same parameter $v_M$, and no realistic spectrum can be achieved. This is the main reason for introducing the additional $N', \bar{N'}$ fields: by dialing the parameter $m_{N'}$ we can separately control the messenger mixing angles, to allow the cancellation in $b_M$, while keeping a positive and comparable size $m_M^2$.

We have calculated the 1-loop messenger ($\phi, \bar{\phi}, N, \bar{N}, N', \bar{N'}$) contributions to $m_D, b_M$ and $m_M^2$, and evaluated them numerically as functions of the parameters of the model. The calculation includes the effects of messenger mixings, and is incorporating terms linear and quadratic in $D$ (using the insertion approximation for the $\phi$ propagator).
Since $\frac{D}{m_{\phi}^2}<1$ we expect the higher order in D contributions to be negligible.

As explained in  Sec.~\ref{withmixing}, to obtain a sufficiently small $b_M$  we need a cancellation between  contributions linear and quadratic in $D$. This is possible because the term linear in $D$ is also  proportional to  $|\psi|^2 -|\bar{\psi}|^2$ (see (\ref{eq:Yuri})), and thus can be further suppressed.

\begin{figure}[H]
\centering{
 \includegraphics[width=.5\textwidth]{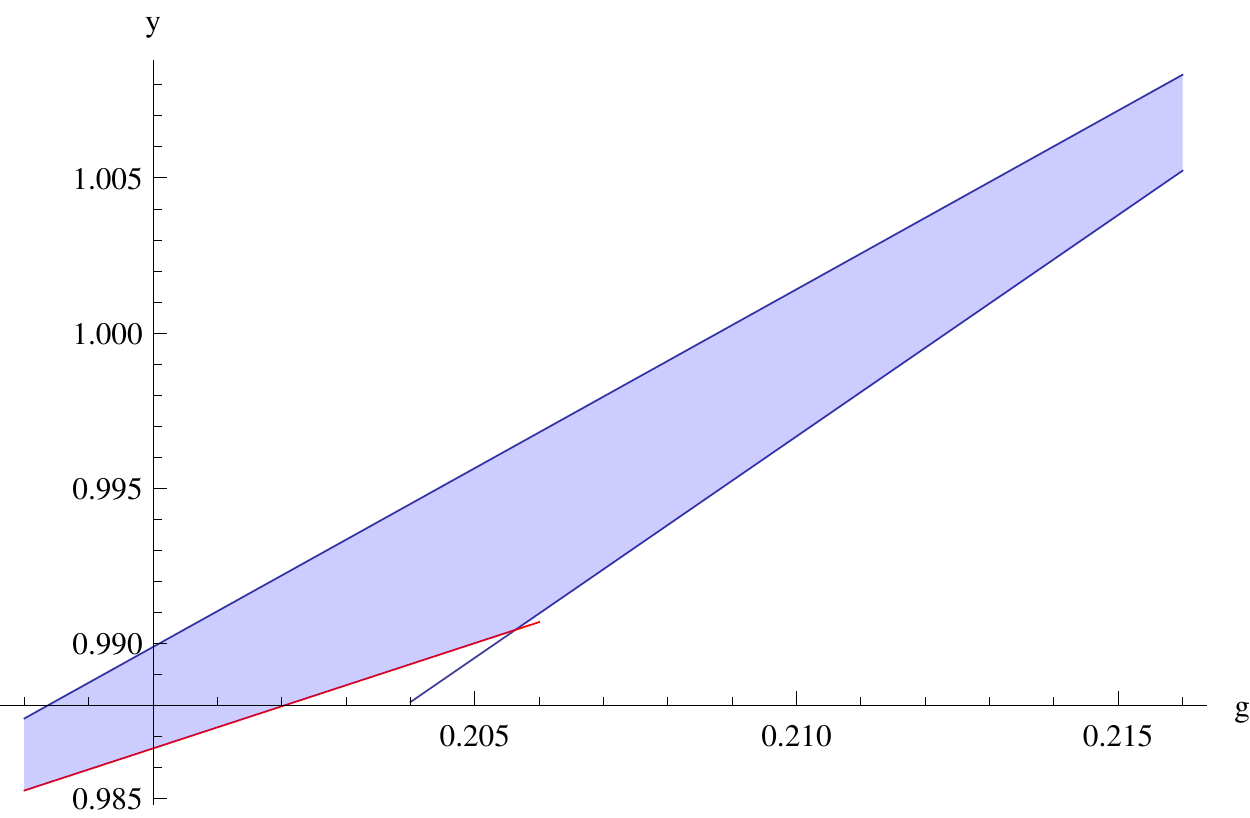}
}
\caption{The phenomenologically acceptable region of the Yukawa couplig $y$ and $U(1)$ gauge coupling g, with $\alpha=1.7$, $h=10h'=5\mu$, $v_M=5\mu$ and $m_{N'}=1.15\mu$. The blue lines indicate the boundaries of the region where all scalar mass squares are positive and the real adjoint is heavier than the gaugino. Points below the red line do not satisfy the bound (\ref{eq:mMbound}) implying negative mass squares for sfermions.}
\label{fig:Yukvsg}
\end{figure}

\begin{figure}[H]
\centering{
 \includegraphics[width=.5\textwidth]{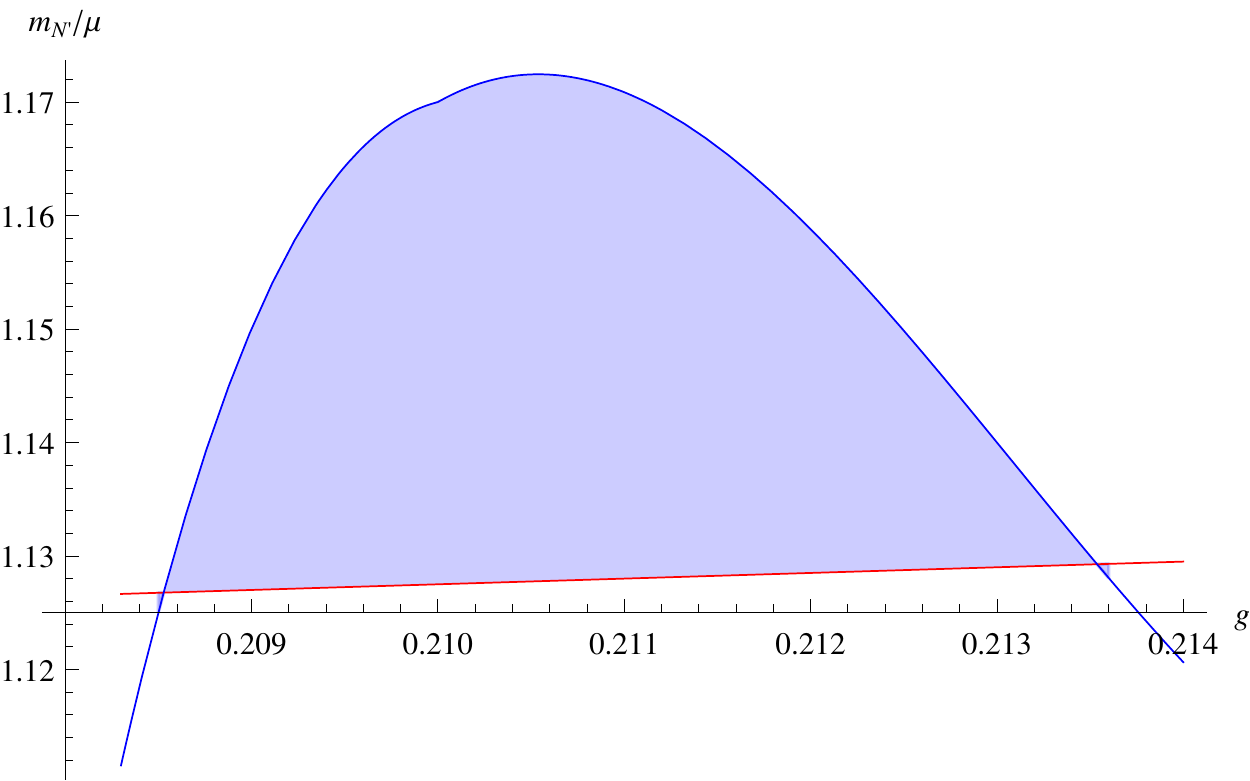}
}
\caption{Same as in Fig.~\ref{fig:Yukvsg}, now as a function of the messenger mixing parameter $m_{N'}$ and $U(1)$ gauge coupling $g$, fixing $y=1$. }
\label{fig:nprimevsg}
\end{figure}

\begin{figure}[H]
\centering{
 \includegraphics[width=.5\textwidth]{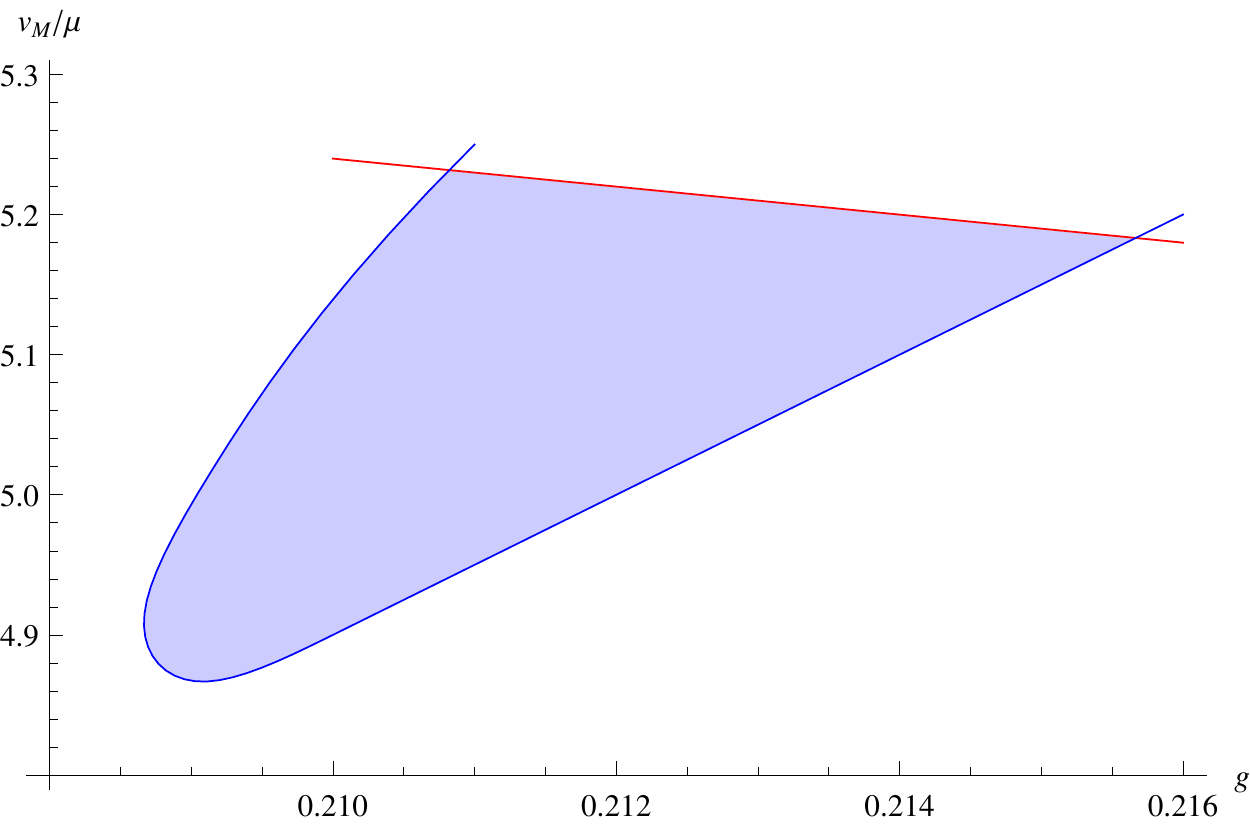}
}
\caption{Same as in Fig.~\ref{fig:Yukvsg}, now as a function of the VEV of Tr$M$, supersymmetric messenger mass, and $U(1)$ gauge coupling $g$, fixing $y=1$.}
\label{fig:trmvsg}
\end{figure}

We did not calculate the two-loop contributions to $m_M^2$ and $b_M$. They are expected to be of the same order as $m_D^2$, and hence relevant for the precise calculation of the spectrum, however they will not qualitatively modify our conclusions. Such corrections will results in shifts of phenomenologically interesting regions without modifying the amount of tuning required to suppress $b_M$.
The region of parameters with phenomenologically acceptable values of the scalar adjoint masses is small: a reflection of the fine-tuning needed in order to solve the $m_D-b_M$ problem.
We present examples of acceptable parameter ranges in Figs.~\ref{fig:Yukvsg}, \ref{fig:nprimevsg} and \ref{fig:trmvsg}. In all cases we find regions with positive scalar mass squares satisfying the bound (\ref{eq:mMbound}), we also require the real adjoint to be heavier than the gaugino for positivity of the MSSM sfermion masses in (\ref{eq:sfermionmass}). 

\begin{figure}[H]
\centering{
 \includegraphics[width=.5\textwidth]{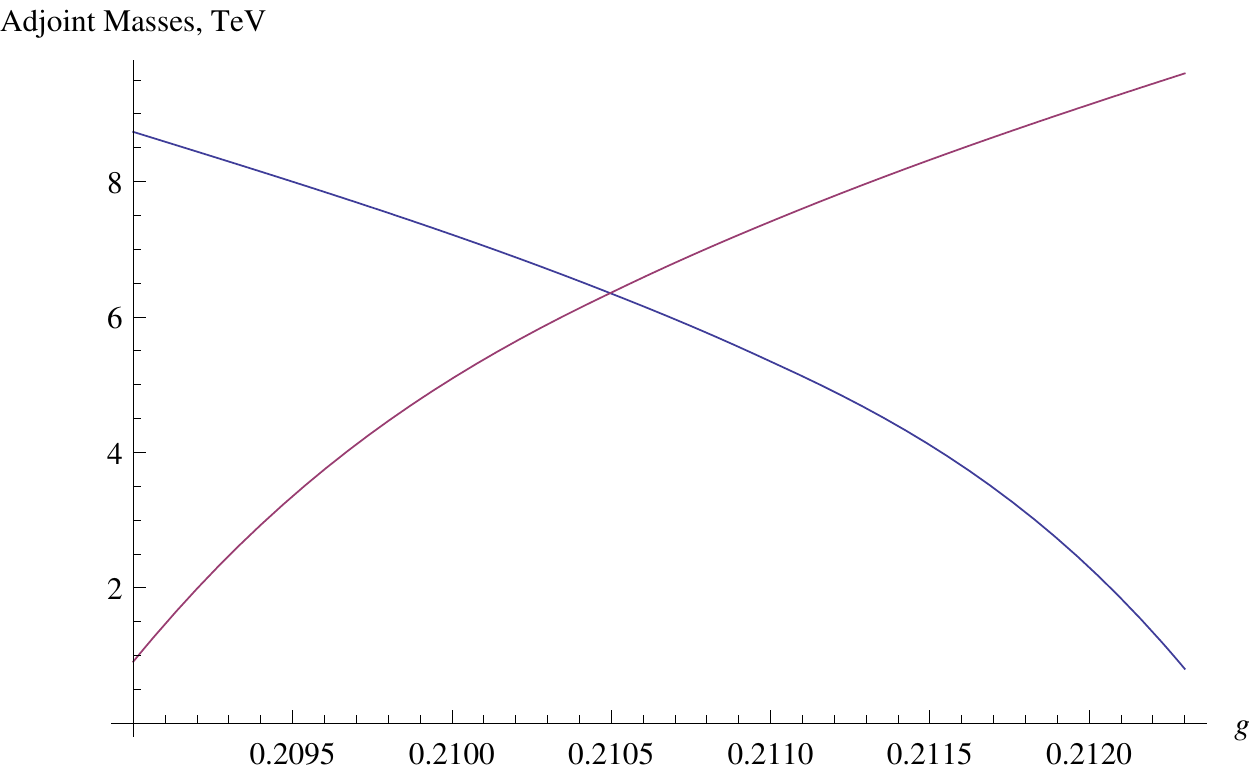}
}
\caption{Scalar adjoint masses (real part (blue) and imaginary part (red)) for the case of 2.5 TeV gluino, as a function of the gauge coupling $g$. The Yukawa is set to $y=1$ and all the other parameters are fixed as in Fig.~\ref{fig:Yukvsg}}
\label{fig:adjmasses}
\end{figure}

In Fig. \ref{fig:adjmasses} we show a plot of adjoint scalar masses as a function of $U(1)$ gauge coupling $g$ and gluino mass set at $2.5~\TeV$. We can see the effect of  a growing $b_M$ through one of the two scalar mass squares becoming negative: $b_M$ is sufficiently cancelled only for a small region of the $U(1)$ coupling parameter $g$, as you move away from the center of the region in either direction $b_M$ will start dominating over the other parameters leading to unphysical spectra.

%%%%%%%%%%%%%%%%%%%%
%%%%%%%%%%%%%%%%%%%%
The ratio of the gaugino masses is given by the ratio of the SM gauge couplings, and to set the overall scale we fix the gluino mass to 2.5 TeV. This splitting in the gaugino masses implies that the masses of the scalar adjoints are also split. In particular the real parts have a $4m_D^2$ contribution to their mass squares. On the other hand, the imaginary parts of the adjoints are degenerate as their masses only come from the loops which respect the $SU(5)$ symmetry.

From Eq.(\ref{eq:sfermionmass}) we see that $\delta_+>m_D$ is also required: otherwise the sfermion mass squares become negative. Thus the real part of the scalar adjoint needs to be heavier than the corresponding gaugino. While this is not a significant constraint on the parameter space, we have included it in the plot of the available parameter space in Figs.~\ref{fig:Yukvsg}, \ref{fig:nprimevsg} and \ref{fig:trmvsg}.

Finally we comment on the off-diagonal chiral superfields of the $SU(5)$ adjoint $M$. While the scalars also get some loop contributions analogous to (\ref{eq:ssoftbadguy}) and  (\ref{eq:Yuri}), the fermionic components remain massless. However, these can be removed by adding explicit superpotential coupling to elementary fields $X$, $Y$, which could give masses as heavy as $\Lambda$.

Ordinary gauge mediated masses, from messenger loops, can be approximately estimated using the result in \cite{Poppitz:1996xw}. The tree-level messenger supertrace vanishes and so the contribution is not logarithmically enhanced but finite, the precise contribution is not known but estimated to be negligible. An F-term SUSY breaking contribution to the messenger masses, which could potentially give a non-vanishing supertrace and hence larger and different gauge-mediation, appears at 1-loop, from $F_X$ and $F_Z$, with two $F$ insertions and heavy $\psi$ and $T$ fields in the loop. We estimated this diagram and found that this contribution is negligible compared to the tree-level D-term contribution, as long as the D-term is not too small compared to the F-terms.

%%%%%%%%%%%%%%%%%%%%
%%%%%%%%%%%%%%%%%%%%
\section{Sketches of Phenomenology\label{secpheno}}
\setcounter{equation}{0}
%%%%%%%%%%%%%%%%%%%%
%%%%%%%%%%%%%%%%%%%%
In this final section we make some brief comments on the basic phenomenology of Dirac gaugino models, applying some of the general results obtained here.

The first comment is that for a pure supersoft mediation model with exact R-symmetry the slepton masses turn out to be very light~\cite{BFGP}, unless the squark masses are raised above 2.5 TeV.
The reason for this is that if (\ref{eq:sfermionmass}) is applied, together with  $m_{Di}^2\propto\alpha_i$ (\ref{eq:gauginoloop}), then we find $m_{\tilde{f}_i}\propto\alpha_i$. This implies that the squark masses are about ten times heavier than the RH sleptons~\footnote{We thank Claudia Frugiuele and Eduardo Pont\'on for emphasizing this point. For a more detailed analysis see~\cite{BFGP}.}. This might actually not be a problem once the RG running of these masses is included. In fact the $m_M^2$ operators, which cause the running in the first place, make the colored fields' masses run faster than the electroweak fields (in particular the righ-handed sleptons do not receive any contributions from the hypercharge ``adjoint" scalars at all) and so it is possible that the squark and slepton masses will not be too separated in the end.

Assuming that there are no large splittings between scalar and fermion in the adjoint sector (as suggested by our discussion of the $m_D-b_M$ problem for the colored particles and assuming the $SU(5)$ global symmetry as in the explicit model presented before), we find that the mass spectrum is of the form presented in Table \ref{spectrum}. We can see that the RH sleptons are well below the LHC (and even LEP) bounds. A simple way to get around this problem is to assume that the pure R-symmetric Dirac gaugino structure only applies to the colored sector, giving rise to Dirac gluinos of order 2.5 TeV, while there could be an additional ordinary Gauge Mediation sector where the messengers are only charged under weak hypercharge, raising all sfermion masses by a few hundred GeV. Another possible way to solve the slepton mass problem is to enhance the log in (\ref{eq:sfermionmass}). This could originate from a heavy scalar $U(1)_Y$ ``adjoint", which is possible in general models (though this does not happen in the dynamical model presented here due to the $SU(5)$ symmetry).

\begin{table}[H]
\centering
\begin{tabular}{|l|r|}
\hline
Squarks &830 GeV	\\
\hline
Left Sleptons&210 GeV	\\
\hline
Right Sleptons & 100 GeV\\
\hline
Gluino&2.5 TeV		\\
\hline
Real Ocet, real Triplet, real scalar singlet&7 TeV, 6 TeV, 5.5 TeV\\
\hline
Imaginary Adjoints &5 TeV	\\
\hline
Winos& 1.3 TeV\\
\hline
Bino&940 GeV\\
\hline
Higgsinos &$\mu_u\sim\mu_d$	\\
\hline
\end{tabular}
\caption{A sample spectrum assuming minimal supersoft mediation with the gluino mass fixed at 2.5 TeV, and using a representative point from the dynamical model presented here for the scalar adjoint spectrum}
\label{spectrum}
\end{table}

Another potentially related general problem with pure R-symmetric Dirac gaugino models is the Higgs sector. An ordinary $\mu$-term would break the R-symmetry, thus it was proposed in \cite{Kribs:2007ac} to extend the Higgs sector by the R-Higgses to an R-symmetric form \cite{Choi:2010an}:
\begin{equation}
W=\mu_u H_u R_u+\mu_d H_d R_d
\end{equation}
with a conventional $B_\mu$ term.\footnote{For an alternative approach to the Higgs sector see~\cite{McCullough}.} 
However, even this R-symmetric form does not predict a sufficiently heavy Higgs mass: in fact the $SU(2)_L$ D-term is suppressed compared to its MSSM value (see \cite{Fox:2002bu}), on the other hand the naturalness of the Higgs potential is improved due to logarithmic corrections being cut off by the Dirac mass. Thus an extension of the Higgs sector to incorporate an NMSSM-type coupling (perhaps originating from the same strong dynamics as in~\cite{compositeMSSM}) or other mechanism to raise the Higgs quartic is essential.
We leave the incorporation of a realistic Higgs sector into a dynamical model to future work.

An interesting phenomenological feature of our model is the existence of the light scalar $SU(3)$ octets, called ``sgluons". The mass of the imaginary part of the sgluons is not constrained and could be lower than the gluino mass. Its pair-production channels are essentially the same as for squarks, enhanced slightly by color factors but reduced overall by the larger mass. It interacts only with gluons and gluinos, and hence its predominant decays are loop processes into $t\bar{t}$ leaving a signature signal of 4 tops. When mixing among the squarks is allowed decays to $t\bar{q}$ and $\bar{t}q$ are also possible, leading to interesting same sign di-lepton final states \cite{Plehn:2008ae}. Their next-to-leading-order interactions for production and decays have been studied in \cite{GoncalvesNetto:2012nt}. Their phenomenology and actual searches are discussed in \cite{Plehn:2008ae, Arnold:2011ra, ATLAS:2012ds, Calvet:2012rk, Idilbi:2010rs, Choi:2008ub, Choi:2009ue}.

Since we have a low scale SUSY breaking scenario, the LSP is always the gravitino. There have been several interesting studies of the phenomenology of the electroweak charge sector of sleptons, neutralinos and charginos \cite{Kribs:2008hq}. However, we expect that once a complete model incorporating a heavy Higgs and sufficiently heavy sleptons is found, its phenomenology will strongly depend on the new particles and interactions.

%%%%%%%%%%%%%%%%%%%%
%%%%%%%%%%%%%%%%%%%%
\section{Conclusions\label{conclusion}}
In this paper we examined models with Dirac gaugino masses and showed that they suffer from the $m_D-b_M$ problem analogous to the $\mu-B_\mu$ problem in the Higgs sector. To better understand the issue we analyzed the effective operators arising in supersoft models and their UV completions. We showed that in models with field-dependent D-terms Dirac gaugino models are never truly supersoft -- non-supersoft operators are always generated at higher order in perturbation theory. Such operators might be non-negligible unless supersymmetry is broken at a very low scale (or the fields Higgsing the $U(1)$ are tuned to be much heavier than both the $U(1)$ breaking and the messenger scales). We also showed that in models with messenger mixing non-supersoft operators are generated already at leading order in the loop expansion. Such terms can be used to solve the $m_D-b_M$ problem of Dirac gauginos implying a tuning of order  $\mathcal{O}(1/(16\pi^2))$.   We constructed a fully dynamical model of supersymmetry breaking in which supersoft SUSY breaking gives the dominant contributions to the soft masses in the gaugino and squark sectors. Finally, while some of the experimental signatures of the model depend on the mechanism generating soft parameters in the Higgs and slepton sector, we pointed out several distinctive phenomenological properties of this class of models.

%%%%%%%%%%%%%%%%%%%%
%%%%%%%%%%%%%%%%%%%%

%%%%%%%%%%%%%%%%%%%%
%%%%%%%%%%%%%%%%%%%%
\appendix
%%%%%%%%%%%%%%%%%%%%
%%%%%%%%%%%%%%%%%%%%

\section*{Appendix}

\section{An Alternative Mechanism for Realistic Adjoint Masses\label{App:discrete}}
\setcounter{equation}{0}

In this appendix we will show a different approach towards obtaining the realistic mass spectrum $m_D^2\sim b_M\sim m_M^2$. As pointed out in~\cite{Fox:2002bu}, there is no symmetry which allows (\ref{eq:ssoft}) while forbidding (\ref{eq:ssoftbadguy}). It is possible, however, to impose a symmetry that forbids both of these operators. If the symmetry is then softly broken, it is possible to choose soft breaking parameters such that the desired relation between soft masses is obtained. As a starting point, let us consider a model with three messenger fields and the superpotential
\begin{equation}
 W=\sum_{k=1}^3\lambda_k \phi_k M\bar \phi_k\,.
\end{equation}
If the coupling constants $\lambda_k$ are related by
\begin{equation}
 \lambda_1=e^{2\pi i/3}\lambda_2=e^{4\pi i/3}\lambda_3=\lambda\,,
\end{equation}
the model possesses a $Z_3$ symmetry 
under which the messenger fields transform as
\begin{equation}
 \phi_k\rightarrow e^{-2\pi i/3}\phi_{k+1},~~~\bar\phi_k\rightarrow\bar\phi_k\,.
\end{equation} 
In this model the one loop contributions to the gaugino and adjoint masses  vanish
since the coefficient of (\ref{eq:ssoft}) is proportional to $\sum_k \lambda_k=0$ while the coefficient of (\ref{eq:ssoftbadguy}) is proportional to $\sum_k\lambda_k^2=0$. We now softly break the symmetry by modifying the coupling constants according to
\begin{equation}
 \lambda_1=(1+2\epsilon_1)\lambda,~~~\lambda_2=(e^{-2\pi i/3}+\epsilon_2)\lambda,~~~\lambda_3=(e^{-4\pi i/3}+\epsilon_3)\lambda\,.
\end{equation}
Generically, 
one finds $m_D^2\sim m_M^2 \sim \mathcal{O}(\frac{\epsilon^2}{(16\pi^2)^2})$ while $b_M \sim \frac{\epsilon}{16\pi^2}$, thus the soft symmetry breaking parameters generically exacerbate $m_D-b_M$ problem, unless $\epsilon \sim \mathcal{O}(1)$. 
However, if one chooses  all soft breaking terms equal $\epsilon_k=\epsilon$ one finds that the leading contribution to all soft mass squares is proportional to $\epsilon^2$. On the other hand, the loop factor suppression of $m_D^2$ and $m_M^2$ compared to $b_M$ remains. 
However, the ability to adjust the three symmetry breaking parameters independently allows us to make all the soft terms comparable and achieve positive eigenvalues for both scalars in the adjoint multiplet. As expected this requires a tuning of order $1/(16\pi^2)$.

%%%%%%%%%%%%%%%%%%%%
%%%%%%%%%%%%%%%%%%%%
\section*{Acknowledgements}
We thank Marco Farina, Claudia Frugiuele, Graham Kribs, Linda Carpenter, Eduardo Pont\'on and Ken Van Tilburg for useful discussions.   C.C. and R.P. are supported in part by the NSF grant PHY-0757868.Y.S. is supported in part by the NSF grant  PHY-1316792. C.C. and Y.S. thank the Galileo Galilei Institute for Theoretical Physics/INFN and C.C. thanks the Aspen Center for Physics for their hospitality while part of this work was completed.
%%%%%%%%%%%%%%%%%%%%
%%%%%%%%%%%%%%%%%%%%


\begin{thebibliography}{99}

\bibitem{AtlasSUSY}
ATLAS Collaboration, 
%``Search for squarks and gluinos with the ATLAS detector in  final states with jets and missing transverse momentum and 20.3 fb$^{-1}$ of $\sqrt{s}=8$ TeV proton-proton collision data",
 ATLAS-CONF-2013-047.


\bibitem{CMSSUSY}
CMS Collaboration, 
%``Search for New Physics in the Multijets and Missing Momentum Final State in Proton-Proton Collisions at 8 TeV", 
CMS-PAS-SUS-13-012.

\bibitem{moreminimal}
 A.~G.~Cohen, D.~B.~Kaplan and A.~E.~Nelson,
  %``The More minimal supersymmetric standard model,''
  Phys.\ Lett.\ B {\bf 388}, 588 (1996)
  [hep-ph/9607394].
  %%CITATION = HEP-PH/9607394;%%
  %498 citations counted in INSPIRE as of 11 Oct 2013


\bibitem{naturalsusy}
 M.~Papucci, J.~T.~Ruderman and A.~Weiler,
  %``Natural SUSY Endures,''
  JHEP {\bf 1209}, 035 (2012)
  [arXiv:1110.6926 [hep-ph]];
  %%CITATION = ARXIV:1110.6926;%%
 Y.~Kats, P.~Meade, M.~Reece and D.~Shih,
  %``The Status of GMSB After 1/fb at the LHC,''
  JHEP {\bf 1202}, 115 (2012)
  [arXiv:1110.6444 [hep-ph]].
  %%CITATION = ARXIV:1110.6444;%%
  
 %\cite{Brust:2011tb}
\bibitem{Brust:2011tb} 
  C.~Brust, A.~Katz, S.~Lawrence and R.~Sundrum,
  %``SUSY, the Third Generation and the LHC,''
  JHEP {\bf 1203}, 103 (2012)
  [arXiv:1110.6670 [hep-ph]].
  %%CITATION = ARXIV:1110.6670;%%
  %143 citations counted in INSPIRE as of 10 Oct 2013

\bibitem{stealth}
 J.~Fan, M.~Reece and J.~T.~Ruderman,
  %``Stealth Supersymmetry,''
  JHEP {\bf 1111}, 012 (2011)
  [arXiv:1105.5135 [hep-ph]];
  %%CITATION = ARXIV:1105.5135;%%
%J.~Fan, M.~Reece and J.~T.~Ruderman,
  %``A Stealth Supersymmetry Sampler,''
  JHEP {\bf 1207}, 196 (2012)
  [arXiv:1201.4875 [hep-ph]];
  %%CITATION = ARXIV:1201.4875;%%
C.~Csaki, L.~Randall and J.~Terning,
  %``Light Stops from Seiberg Duality,''
  Phys.\ Rev.\ D {\bf 86}, 075009 (2012)
  [arXiv:1201.1293 [hep-ph]];
  %%CITATION = ARXIV:1201.1293;%%

\bibitem{RPV}
 C.~Csaki, Y.~Grossman and B.~Heidenreich,
  %``MFV SUSY: A Natural Theory for R-Parity Violation,''
  Phys.\ Rev.\ D {\bf 85}, 095009 (2012)
  [arXiv:1111.1239 [hep-ph]];
  %%CITATION = ARXIV:1111.1239;%%
C.~Brust, A.~Katz and R.~Sundrum,
  %``SUSY Stops at a Bump,''
  JHEP {\bf 1208}, 059 (2012)
  [arXiv:1206.2353 [hep-ph]];
  %%CITATION = ARXIV:1206.2353;%%
 P.~W.~Graham, D.~E.~Kaplan, S.~Rajendran and P.~Saraswat,
  %``Displaced Supersymmetry,''
  JHEP {\bf 1207}, 149 (2012)
  [arXiv:1204.6038 [hep-ph]].
  %%CITATION = ARXIV:1204.6038;%%
 P.~Fileviez Perez and S.~Spinner,
  %``The Minimal Theory for R-parity Violation at the LHC,''
  JHEP {\bf 1204}, 118 (2012)
  [arXiv:1201.5923 [hep-ph]];
  %%CITATION = ARXIV:1201.5923;%%
J.~A.~Evans and Y.~Kats,
  %``LHC Coverage of RPV MSSM with Light Stops,''
  arXiv:1209.0764 [hep-ph];
  %%CITATION = ARXIV:1209.0764;%%
J.~T.~Ruderman, T.~R.~Slatyer and N.~Weiner,
  %``A Collective Breaking of R-Parity,''
  arXiv:1207.5787 [hep-ph];
  %%CITATION = ARXIV:1207.5787;%%
Z.~Han, A.~Katz, M.~Son and B.~Tweedie,
  %``Boosting Searches for Natural SUSY with RPV via Gluino Cascades,''
  arXiv:1211.4025 [hep-ph];
  %%CITATION = ARXIV:1211.4025;%%
 R.~Franceschini and R.~Torre,
  %``RPV stops bump off the background,''
  arXiv:1212.3622 [hep-ph];
  %%CITATION = ARXIV:1212.3622;%%
 C.~Csaki, E.~Kuflik and T.~Volansky,
  %``Dynamical R-Parity Violation,''
  arXiv:1309.5957 [hep-ph].
  %%CITATION = ARXIV:1309.5957;%%
  %1 citations counted in INSPIRE as of 10 Oct 2013

%\cite{Fox:2002bu}
\bibitem{Fox:2002bu} 
  P.~J.~Fox, A.~E.~Nelson and N.~Weiner,
  %``Dirac gaugino masses and supersoft supersymmetry breaking,''
  JHEP {\bf 0208}, 035 (2002)
  [hep-ph/0206096].
  %%CITATION = HEP-PH/0206096;%%
  %133 citations counted in INSPIRE as of 27 May 2013
  
\bibitem{Fayet}
 P.~Fayet,
  %``Massive Gluinos,''
  Phys.\ Lett.\ B {\bf 78}, 417 (1978).
  %%CITATION = PHLTA,B78,417;%%
  %102 citations counted in INSPIRE as of 31 Oct 2013


%\cite{Hall:1990hq}
\bibitem{Hall:1990hq} 
  L.~J.~Hall and L.~Randall,
  %``U(1)-R symmetric supersymmetry,''
  Nucl.\ Phys.\ B {\bf 352}, 289 (1991).
  %%CITATION = NUPHA,B352,289;%%
  %104 citations counted in INSPIRE as of 10 Oct 2013)  
  
 
  
 %\cite{Kribs:2012gx}
\bibitem{Kribs:2012gx} 
  G.~D.~Kribs and A.~Martin,
  %``Supersoft Supersymmetry is Super-Safe,''
  Phys.\ Rev.\ D {\bf 85}, 115014 (2012)
  [arXiv:1203.4821 [hep-ph]];
  %%CITATION = ARXIV:1203.4821;%%
  %27 citations counted in INSPIRE as of 12 Jun 2013
  %\cite{Kribs:2013oda}
%\bibitem{Kribs:2013oda} 
%  G.~D.~Kribs and A.~Martin,
  %``Dirac Gauginos in Supersymmetry -- Suppressed Jets + MET Signals: A Snowmass Whitepaper,''
  arXiv:1308.3468 [hep-ph].
  %%CITATION = ARXIV:1308.3468;%%
  %1 citations counted in INSPIRE as of 10 Oct 2013
  
  %\cite{Kribs:2007ac}
\bibitem{Kribs:2007ac} 
  G.~D.~Kribs, E.~Poppitz and N.~Weiner,
  %``Flavor in supersymmetry with an extended R-symmetry,''
  Phys.\ Rev.\ D {\bf 78}, 055010 (2008)
  [arXiv:0712.2039 [hep-ph]].
  %%CITATION = ARXIV:0712.2039;%%
  %100 citations counted in INSPIRE as of 12 Jun 2013
  
\bibitem{Goodsell}
K.~Benakli and M.~D.~Goodsell,
  %``Dirac Gauginos in General Gauge Mediation,''
  Nucl.\ Phys.\ B {\bf 816}, 185 (2009)
  [arXiv:0811.4409 [hep-ph]];
  %%CITATION = ARXIV:0811.4409;%%
  %56 citations counted in INSPIRE as of 10 Oct 2013;
% K.~Benakli and M.~D.~Goodsell,
  %``Dirac Gauginos, Gauge Mediation and Unification,''
  Nucl.\ Phys.\ B {\bf 840}, 1 (2010)
  [arXiv:1003.4957 [hep-ph]];
  %%CITATION = ARXIV:1003.4957;%%
  %34 citations counted in INSPIRE as of 28 Jun 2013;

  %\cite{Benakli}
\bibitem{Benakli} 
 K.~Benakli and M.~D.~Goodsell,
  %``Dirac Gauginos and Kinetic Mixing,''
  Nucl.\ Phys.\ B {\bf 830}, 315 (2010)
  [arXiv:0909.0017 [hep-ph]];
  %%CITATION = ARXIV:0909.0017;%%
  %37 citations counted in INSPIRE as of 10 Oct 2013;
    K.~Benakli,
  %``Dirac Gauginos: A User Manual,''
  Fortsch.\ Phys.\  {\bf 59}, 1079 (2011)
  [arXiv:1106.1649 [hep-ph]];
  %%CITATION = ARXIV:1106.1649;%%
  %17 citations counted in INSPIRE as of 10 Oct 2013;
   K.~Benakli, M.~D.~Goodsell and A.~-K.~Maier,
  %``Generating mu and Bmu in models with Dirac Gauginos,''
  Nucl.\ Phys.\ B {\bf 851}, 445 (2011)
  [arXiv:1104.2695 [hep-ph]];
  %%CITATION = ARXIV:1104.2695;%%
  %17 citations counted in INSPIRE as of 10 Oct 2013;
  K.~Benakli, M.~D.~Goodsell and F.~Staub,
  %``Dirac Gauginos and the 125 GeV Higgs,''
  JHEP {\bf 1306}, 073 (2013)
  [arXiv:1211.0552 [hep-ph]].
  %%CITATION = ARXIV:1211.0552;%%
  %18 citations counted in INSPIRE as of 10 Oct 2013;
  
  %\cite{Belanger:2009wf}
\bibitem{Belanger:2009wf} 
  G.~Belanger, K.~Benakli, M.~Goodsell, C.~Moura and A.~Pukhov,
  %``Dark Matter with Dirac and Majorana Gaugino Masses,''
  JCAP {\bf 0908}, 027 (2009)
  [arXiv:0905.1043 [hep-ph]].
  %%CITATION = ARXIV:0905.1043;%%
  %34 citations counted in INSPIRE as of 10 Oct 2013
  

%\cite{Amigo:2008rc}
\bibitem{Amigo:2008rc} 
  S.~D.~L.~Amigo, A.~E.~Blechman, P.~J.~Fox and E.~Poppitz,
  %``R-symmetric gauge mediation,''
  JHEP {\bf 0901}, 018 (2009)
  [arXiv:0809.1112 [hep-ph]].
  %%CITATION = ARXIV:0809.1112;%%
  %42 citations counted in INSPIRE as of 28 Jun 2013
  
  %\cite{Abel:2011dc}
\bibitem{Abel:2011dc} 
  S.~Abel and M.~Goodsell,
  %``Easy Dirac Gauginos,''
  JHEP {\bf 1106}, 064 (2011)
  [arXiv:1102.0014 [hep-th]].
  %%CITATION = ARXIV:1102.0014;%%
  %24 citations counted in INSPIRE as of 28 Jun 2013
  %\cite{Harnik:2008uu}
  
\bibitem{Harnik:2008uu} 
  R.~Harnik and G.~D.~Kribs,
  %``An Effective Theory of Dirac Dark Matter,''
  Phys.\ Rev.\ D {\bf 79}, 095007 (2009)
  [arXiv:0810.5557 [hep-ph]].
  %%CITATION = ARXIV:0810.5557;%%
  %110 citations counted in INSPIRE as of 10 Oct 2013
  %\cite{Hsieh:2007wq}
\bibitem{Hsieh:2007wq} 
  K.~Hsieh,
  %``Pseudo-Dirac bino dark matter,''
  Phys.\ Rev.\ D {\bf 77}, 015004 (2008)
  [arXiv:0708.3970 [hep-ph]].
  %%CITATION = ARXIV:0708.3970;%%
  %31 citations counted in INSPIRE as of 10 Oct 2013
  
  %\cite{Blechman:2009if}
\bibitem{Blechman:2009if} 
  A.~E.~Blechman,
  %``R-symmetric Gauge Mediation and the Minimal R-Symmetric Supersymmetric Standard Model,''
  Mod.\ Phys.\ Lett.\ A {\bf 24}, 633 (2009)
  [arXiv:0903.2822 [hep-ph]].
  %%CITATION = ARXIV:0903.2822;%%
  %20 citations counted in INSPIRE as of 10 Oct 2013

\bibitem{McCullough}
R.~Davies, J.~March-Russell and M.~McCullough,
  %``A Supersymmetric One Higgs Doublet Model,''
  JHEP {\bf 1104}, 108 (2011)
  [arXiv:1103.1647 [hep-ph]].
  %%CITATION = ARXIV:1103.1647;%%
  %35 citations counted in INSPIRE as of 31 Oct 2013

%\cite{Kribs:2010md}
\bibitem{Kribs:2010md} 
  G.~D.~Kribs, T.~Okui and T.~S.~Roy,
  %``Viable Gravity-Mediated Supersymmetry Breaking,''
  Phys.\ Rev.\ D {\bf 82}, 115010 (2010)
  [arXiv:1008.1798 [hep-ph]].
  %%CITATION = ARXIV:1008.1798;%%
  %19 citations counted in INSPIRE as of 01 Nov 2013

%\cite{dvali}
\bibitem{dvali}
G.~R.~Dvali, G.~F.~Giudice and A.~Pomarol,
  %``The Mu problem in theories with gauge mediated supersymmetry breaking,''
  Nucl.\ Phys.\ B {\bf 478}, 31 (1996)
  [hep-ph/9603238];
  %%CITATION = HEP-PH/9603238;%%
  %231 citations counted in INSPIRE as of 10 Oct 2013
  
\bibitem{muBmu}
 P.~Langacker, N.~Polonsky and J.~Wang,
  %``A Low-energy solution to the mu problem in gauge mediation,''
  Phys.\ Rev.\ D {\bf 60}, 115005 (1999)
  [hep-ph/9905252].
  %%CITATION = HEP-PH/9905252;%%
  %49 citations counted in INSPIRE as of 10 Oct 2013
L.~J.~Hall, Y.~Nomura and A.~Pierce,
  %``R symmetry and the mu problem,''
  Phys.\ Lett.\ B {\bf 538}, 359 (2002)
  [hep-ph/0204062];
  %%CITATION = HEP-PH/0204062;%%
  %31 citations counted in INSPIRE as of 10 Oct 2013
T.~S.~Roy and M.~Schmaltz,
  %``Hidden solution to the mu/Bmu problem in gauge mediation,''
  Phys.\ Rev.\ D {\bf 77}, 095008 (2008)
  [arXiv:0708.3593 [hep-ph]];
  %%CITATION = ARXIV:0708.3593;%%
  %66 citations counted in INSPIRE as of 10 Oct 2013
H.~Murayama, Y.~Nomura and D.~Poland,
  %``More visible effects of the hidden sector,''
  Phys.\ Rev.\ D {\bf 77}, 015005 (2008)
  [arXiv:0709.0775 [hep-ph]].
  %%CITATION = ARXIV:0709.0775;%%
  %75 citations counted in INSPIRE as of 10 Oct 2013
 G.~F.~Giudice, H.~D.~Kim and R.~Rattazzi,
  %``Natural mu and B mu in gauge mediation,''
  Phys.\ Lett.\ B {\bf 660}, 545 (2008)
  [arXiv:0711.4448 [hep-ph]];
  %%CITATION = ARXIV:0711.4448;%%
  %52 citations counted in INSPIRE as of 10 Oct 2013
C.~Csaki, A.~Falkowski, Y.~Nomura and T.~Volansky,
  %``New Approach to the mu-Bmu Problem of Gauge-Mediated Supersymmetry Breaking,''
  Phys.\ Rev.\ Lett.\  {\bf 102}, 111801 (2009)
  [arXiv:0809.4492 [hep-ph]].
  %%CITATION = ARXIV:0809.4492;%%
  %49 citations counted in INSPIRE as of 10 Oct 2013
  
  \bibitem{Stanfordstudents}
  A.~Arvanitaki, M.~Baryakhtar, X.~Huang, K.~Van Tilburg and G.~Villadoro,
  %``The Last Vestiges of Naturalness,''
  arXiv:1309.3568 [hep-ph].
  %%CITATION = ARXIV:1309.3568;%%
  %1 citations counted in INSPIRE as of 10 Oct 2013
  
  %\cite{Goodsell:2012fm}
\bibitem{Goodsell:2012fm} 
  M.~D.~Goodsell,
  %``Two-loop RGEs with Dirac gaugino masses,''
  JHEP {\bf 1301}, 066 (2013)
  [arXiv:1206.6697 [hep-ph]].
  %%CITATION = ARXIV:1206.6697;%%
  %11 citations counted in INSPIRE as of 10 Oct 2013
  
  \bibitem{Martin:1993zk} 
  S.~P.~Martin and M.~T.~Vaughn,
  %``Two loop renormalization group equations for soft supersymmetry breaking couplings,''
  Phys.\ Rev.\ D {\bf 50}, 2282 (1994)
  [Erratum-ibid.\ D {\bf 78}, 039903 (2008)]
  [hep-ph/9311340].
  %%CITATION = HEP-PH/9311340;%%
  %566 citations counted in INSPIRE as of 05 Sep 2013
  
   %\cite{Poppitz:1996xw}
\bibitem{Poppitz:1996xw} 
  E.~Poppitz and S.~P.~Trivedi,
  %``Some remarks on gauge mediated supersymmetry breaking,''
  Phys.\ Lett.\ B {\bf 401}, 38 (1997)
  [hep-ph/9703246].
  %%CITATION = HEP-PH/9703246;%%
  %92 citations counted in INSPIRE as of 25 Jun 2013
  
  %\cite{Carpenter:2010as}
\bibitem{Carpenter:2010as} 
  L.~M.~Carpenter,
  %``Dirac Gauginos, Negative Supertraces and Gauge Mediation,''
  JHEP {\bf 1209}, 102 (2012)
  [arXiv:1007.0017 [hep-th]].
  %%CITATION = ARXIV:1007.0017;%%
  %14 citations counted in INSPIRE as of 27 May 2013

%\cite{Dine:1996xk}
\bibitem{Dine:1996xk} 
  M.~Dine, Y.~Nir and Y.~Shirman,
  %``Variations on minimal gauge mediated supersymmetry breaking,''
  Phys.\ Rev.\ D {\bf 55}, 1501 (1997)
  [hep-ph/9607397].
  %%CITATION = HEP-PH/9607397;%%
  %189 citations counted in INSPIRE as of 13 Oct 2013  

%\cite{Giudice:1997ni}
\bibitem{Giudice:1997ni} 
  G.~F.~Giudice and R.~Rattazzi,
  %``Extracting supersymmetry breaking effects from wave function renormalization,''
  Nucl.\ Phys.\ B {\bf 511}, 25 (1998)
  [hep-ph/9706540].
  %%CITATION = HEP-PH/9706540;%%
  %241 citations counted in INSPIRE as of 07 Oct 2013
  

  
\bibitem{Intriligator:2010be} 
  K.~Intriligator and M.~Sudano,
  %``General Gauge Mediation with Gauge Messengers,''
  JHEP {\bf 1006}, 047 (2010)
  [arXiv:1001.5443 [hep-ph]].
  %%CITATION = ARXIV:1001.5443;%%
  %31 citations counted in INSPIRE as of 14 Oct 2013

%\cite{Seiberg:1994bz}
\bibitem{Seiberg:1994bz} 
  N.~Seiberg,
  %``Exact results on the space of vacua of four-dimensional SUSY gauge theories,''
  Phys.\ Rev.\ D {\bf 49}, 6857 (1994)
  [hep-th/9402044];
  %%CITATION = HEP-TH/9402044;%%
  %586 citations counted in INSPIRE as of 10 Oct 2013
  %``Electric - magnetic duality in supersymmetric nonAbelian gauge theories,''
  Nucl.\ Phys.\ B {\bf 435}, 129 (1995)
  [hep-th/9411149].
  %%CITATION = HEP-TH/9411149;%%
  %1090 citations counted in INSPIRE as of 29 May 2013
  %``The Power of duality: Exact results in 4-D SUSY field theory,''
  Int.\ J.\ Mod.\ Phys.\ A {\bf 16}, 4365 (2001)
  [hep-th/9506077].
  %%CITATION = HEP-TH/9506077;%%
  %62 citations counted in INSPIRE as of 29 May 2013


  
%\cite{Csaki:2006wi}
\bibitem{Csaki:2006wi} 
  C.~Csaki, Y.~Shirman and J.~Terning,
  %``A Simple Model of Low-scale Direct Gauge Mediation,''
  JHEP {\bf 0705}, 099 (2007)
  [hep-ph/0612241].
  %%CITATION = HEP-PH/0612241;%%
  %109 citations counted in INSPIRE as of 27 May 2013
  
  %\cite{Csaki:1996sm}
\bibitem{Csaki:1996sm} 
  C.~Csaki, M.~Schmaltz and W.~Skiba,
  %``A Systematic approach to confinement in N=1 supersymmetric gauge theories,''
  Phys.\ Rev.\ Lett.\  {\bf 78}, 799 (1997)
  [hep-th/9610139];
  %%CITATION = HEP-TH/9610139;%%
  %47 citations counted in INSPIRE as of 10 Oct 2013
  %``Confinement in N=1 SUSY gauge theories and model building tools,''
  Phys.\ Rev.\ D {\bf 55}, 7840 (1997)
  [hep-th/9612207].
  %%CITATION = HEP-TH/9612207;%%
  %89 citations counted in INSPIRE as of 10 Oct 2013
  
    %\cite{Intriligator:2006dd}
\bibitem{Intriligator:2006dd} 
  K.~A.~Intriligator, N.~Seiberg and D.~Shih,
  %``Dynamical SUSY breaking in meta-stable vacua,''
  JHEP {\bf 0604}, 021 (2006)
  [hep-th/0602239].
  %%CITATION = HEP-TH/0602239;%%
  %465 citations counted in INSPIRE as of 27 May 2013
  
  %\cite{Dumitrescu:2010ca}
\bibitem{Dumitrescu:2010ca} 
  T.~T.~Dumitrescu, Z.~Komargodski and M.~Sudano,
  %``Global Symmetries and D-Terms in Supersymmetric Field Theories,''
  JHEP {\bf 1011}, 052 (2010)
  [arXiv:1007.5352 [hep-th]].
  %%CITATION = ARXIV:1007.5352;%%
  %21 citations counted in INSPIRE as of 17 Jun 2013

%\cite{BFGP}
\bibitem{BFGP}
E.~Bertuzzo, C.~Frugiuele, T.~Gr\'egoire, and E.~Pont\'on, to appear
  
  %\cite{Choi:2010an}
\bibitem{Choi:2010an} 
  S.~Y.~Choi, D.~Choudhury, A.~Freitas, J.~Kalinowski and P.~M.~Zerwas,
  %``The Extended Higgs System in $R$-symmetric Supersymmetry Theories,''
  Phys.\ Lett.\ B {\bf 697}, 215 (2011)
  [Erratum-ibid.\ B {\bf 698}, 457 (2011)]
  [arXiv:1012.2688 [hep-ph]].
  %%CITATION = ARXIV:1012.2688;%%
  %20 citations counted in INSPIRE as of 10 Oct 2013
  

%\cite{compositeMSSM}
  \bibitem{compositeMSSM}
C.~Csaki, Y.~Shirman and J.~Terning,
  %``A Seiberg Dual for the MSSM: Partially Composite W and Z,''
  Phys.\ Rev.\ D {\bf 84}, 095011 (2011)
  [arXiv:1106.3074 [hep-ph]].
  %%CITATION = ARXIV:1106.3074;%%
  %20 citations counted in INSPIRE as of 10 Oct 2013

 %\cite{Plehn:2008ae}
\bibitem{Plehn:2008ae} 
  T.~Plehn and T.~M.~P.~Tait,
  %``Seeking Sgluons,''
  J.\ Phys.\ G {\bf 36}, 075001 (2009)
  [arXiv:0810.3919 [hep-ph]].
  %%CITATION = ARXIV:0810.3919;%%
  %73 citations counted in INSPIRE as of 27 Jun 2013
  
   %\cite{GoncalvesNetto:2012nt}
\bibitem{GoncalvesNetto:2012nt} 
  D.~Goncalves-Netto, D.~Lopez-Val, K.~Mawatari, T.~Plehn and I.~Wigmore,
  %``Sgluon Pair Production to Next-to-Leading Order,''
  Phys.\ Rev.\ D {\bf 85}, 114024 (2012)
  [arXiv:1203.6358 [hep-ph]].
  %%CITATION = ARXIV:1203.6358;%%
  %10 citations counted in INSPIRE as of 27 Jun 2013


%\cite{Calvet:2012rk}
\bibitem{Calvet:2012rk} 
  S.~Calvet, B.~Fuks, P.~Gris and L.~Valery,
  %``Searching for sgluons in multitop events at a center-of-mass energy of 8 TeV,''
  JHEP {\bf 1304}, 043 (2013)
  [arXiv:1212.3360 [hep-ph]].
  %%CITATION = ARXIV:1212.3360;%%
  
  %\cite{ATLAS:2012ds}
\bibitem{ATLAS:2012ds} 
  G.~Aad {\it et al.}  [ATLAS Collaboration],
  %``Search for pair-produced massive coloured scalars in four-jet final states with the ATLAS detector in proton-proton collisions at $\sqrt{s}=7$ TeV,''
  Eur.\ Phys.\ J.\ C {\bf 73}, 2263 (2013)
  [arXiv:1210.4826 [hep-ex]].
  %%CITATION = ARXIV:1210.4826;%%
  %18 citations counted in INSPIRE as of 27 Jun 2013
  
%\cite{Arnold:2011ra}
\bibitem{Arnold:2011ra} 
  J.~M.~Arnold and B.~Fornal,
  %``Color octet scalars and high pT four-jet events at LHC,''
  Phys.\ Rev.\ D {\bf 85}, 055020 (2012)
  [arXiv:1112.0003 [hep-ph]].
  %%CITATION = ARXIV:1112.0003;%%
  %9 citations counted in INSPIRE as of 27 Jun 2013
  
  %\cite{Idilbi:2010rs}
\bibitem{Idilbi:2010rs} 
  A.~Idilbi, C.~Kim and T.~Mehen,
  %``Pair Production of Color-Octet Scalars at the LHC,''
  Phys.\ Rev.\ D {\bf 82}, 075017 (2010)
  [arXiv:1007.0865 [hep-ph]].
  %%CITATION = ARXIV:1007.0865;%%
  %15 citations counted in INSPIRE as of 27 Jun 2013
  
%\cite{Choi:2008ub}
\bibitem{Choi:2008ub} 
  S.~Y.~Choi, M.~Drees, J.~Kalinowski, J.~M.~Kim, E.~Popenda and P.~M.~Zerwas,
  %``Color-Octet Scalars of N=2 Supersymmetry at the LHC,''
  Phys.\ Lett.\ B {\bf 672}, 246 (2009)
  [arXiv:0812.3586 [hep-ph]].
  %%CITATION = ARXIV:0812.3586;%%
  %51 citations counted in INSPIRE as of 27 Jun 2013

%\cite{Choi:2009ue}
\bibitem{Choi:2009ue} 
  S.~Y.~Choi, J.~Kalinowski, J.~M.~Kim and E.~Popenda,
  %``Scalar gluons and Dirac gluinos at the LHC,''
  Acta Phys.\ Polon.\ B {\bf 40}, 2913 (2009)
  [arXiv:0911.1951 [hep-ph]].
  %%CITATION = ARXIV:0911.1951;%%
  %7 citations counted in INSPIRE as of 27 Jun 2013
  
  %\cite{Kribs:2008hq}
\bibitem{Kribs:2008hq} 
  G.~D.~Kribs, A.~Martin and T.~S.~Roy,
  %``Supersymmetry with a Chargino NLSP and Gravitino LSP,''
  JHEP {\bf 0901}, 023 (2009)
  [arXiv:0807.4936 [hep-ph]].
  %%CITATION = ARXIV:0807.4936;%%
  %31 citations counted in INSPIRE as of 25 Jun 2013


\end{thebibliography}
\end{document}